\documentclass[fleqn,usenatbib]{mnras}

\usepackage{newtxtext,newtxmath}

\usepackage[T1]{fontenc}
\usepackage{ae,aecompl}

\usepackage{graphicx}	
\usepackage{amsmath}	
\usepackage{breqn}
\usepackage{xcolor}
\usepackage{cases}
\usepackage{enumerate}
\usepackage{enumitem}
\usepackage{threeparttable}
\usepackage[multidot]{grffile}
\usepackage{multirow}
\usepackage[authoryear]{natbib}
\bibpunct{(}{)}{;}{a}{}{,}

\newcommand{\BlueTides }{\textsc{BlueTides} }
\newcommand{\BlueTidesns}{\textsc{BlueTides}}

\usepackage{makecell}

\title[The \BlueTides Mock Image Catalogue]{The \BlueTides Mock Image Catalogue: Simulated observations of high-redshift galaxies and predictions for \textit{JWST} imaging surveys}

\author[M. A. Marshall et al.]{Madeline A. Marshall$^{1,2}$\thanks{E-mail: Madeline.Marshall@nrc-cnrc.gc.ca (MAM)}, Katelyn Watts$^{1,3}$, Stephen Wilkins$^{4}$, Tiziana Di Matteo$^{5}$, Jussi K. Kuusisto$^4$,  \newauthor
William J. Roper$^4$, Aswin P. Vijayan$^{6,7,4}$, Yueying Ni$^{5}$, Yu Feng$^{8}$,  Rupert A.C. Croft$^5$
\\
$^{1}$ National Research Council of Canada, Herzberg Astronomy \& Astrophysics Research Centre, 5071 West Saanich Road, Victoria, BC V9E 2E7, Canada\\
$^{2}$ ARC Centre of Excellence for All Sky Astrophysics in 3 Dimensions (ASTRO 3D), Australia\\
$^{3}$ Department of Physics and Astronomy, University of Waterloo, 200 University Avenue West, Waterloo, Ontario, N2L 3G1, Canada\\
$^{4}$ Astronomy Centre, Department of Physics and Astronomy, University of Sussex, Brighton, BN1 9QH, UK\\
$^{5}$ McWilliams Center for Cosmology, Department of Physics, Carnegie Mellon University, Pittsburgh, PA 15213, USA \\
$^{6}$ Cosmic Dawn Center (DAWN) \\
$^{7}$ DTU-Space, Technical University of Denmark, Elektrovej 327, DK-2800 Kgs. Lyngby, Denmark \\
$^{8}$ Berkeley Center for Cosmological Physics and Department of Physics, University of California, Berkeley, CA 94720, USA \\}
\date{Accepted 2022 July 22. Received 2022 July 21; in original form 2022 June 13}

\pubyear{2022}

\begin{document}

\label{firstpage}
\pagerange{\pageref{firstpage}--\pageref{lastpage}}
\maketitle

\begin{abstract}
We present a mock image catalogue of $\sim$100,000 $M_{UV}\simeq-22.5$ to $-19.6$ mag galaxies at $z=7$--12 from the \BlueTides cosmological simulation. We create mock images of each galaxy with the \textit{James Webb} (\textit{JWST}), \textit{Hubble}, \textit{Roman}, and \textit{Euclid Space Telescopes}, as well as Subaru, and VISTA, with a range of near- and mid-infrared filters. 
We perform photometry on the mock images to estimate the success of these instruments for detecting high-$z$ galaxies.
We predict that \textit{JWST} will have unprecedented power in detecting high-$z$ galaxies, with a 95\% completeness limit at least 2.5 magnitudes fainter than VISTA and Subaru, 1.1 magnitudes fainter than \textit{Hubble}, and 0.9 magnitudes fainter than \textit{Roman}, for the same wavelength and exposure time.
Focusing on \textit{JWST}, we consider a range of exposure times and filters, and find that the NIRCam F356W and F277W filters will detect the faintest galaxies, with 95\% completeness at $m\simeq27.4$ mag in 10ks exposures. 
We also predict the number of high-$z$ galaxies that will be discovered by upcoming \textit{JWST} imaging surveys.
We predict that the COSMOS-Web survey will detect $\sim$1000 $M_{\textrm{1500\AA}}<-20.1$ mag galaxies at $6.5<z<7.5$, by virtue of its large survey area. JADES-Medium will detect almost $100\%$ of $M_{\textrm{1500\AA}}\lesssim-20$ mag galaxies at $z<8.5$ due to its significant depth, however with its smaller survey area it will detect only $\sim$100 of these galaxies at $6.5<z<7.5$.
Cosmic variance results in a large range in the number of predicted galaxies each survey will detect, which is more evident in smaller surveys such as CEERS and the PEARLS NEP and GOODS-S fields.
\end{abstract}
\begin{keywords}
galaxies: evolution -- galaxies: high-redshift -- galaxies: formation.
\end{keywords}


\section{Introduction}

The detection of the most distant, high-redshift galaxies is a key goal for upcoming surveys with new telescopes. To date, the highest-redshift spectroscopically confirmed galaxy was found at $z=11$ \citep{Oesch2016,Jiang2020a}, with potential candidates recently discovered at $z\sim13$ \citep{Harikane2022}. Finding these most distant objects, and obtaining a statistical sample of fainter, more typical high-$z$ ($z\gtrsim7$) galaxies, provides vital insights into galaxy formation and growth in the early Universe. These sources are also expected to be the primary driver of the reionization of the Universe, and so understanding their population provides information on the mechanisms and timelines of cosmic reionization. 

The majority of high-$z$ sources so far have been discovered by medium and deep surveys with the \textit{Hubble Space Telescope} (\textit{HST}), such as the Hubble Deep Field \citep{Williams1996}, Hubble Ultra Deep Field \citep[HUDF;][]{beckwith_2006,koekemoer_2013}, 
Great Observatories Origins Deep Survey  \citep[GOODS;][]{Giavalisco2004}, Cosmic Evolution Survey \citep[COSMOS;][]{Scoville2007,Koekemoer2007}, Wide Field Camera 3 (WFC3) Early Release Science (ERS) program \citep{Windhorst2011}, Cosmic Assembly Near-infrared Deep Extragalactic Legacy Survey \citep[CANDELS;][]{grogin_2011}, and untargeted pure-parallel surveys such as HIPPIES \citep{yan_2011} and BoRG \citep{Trenti2011}.
One key strategy is to search for magnified high-$z$ galaxies via gravitational lensing from a foreground galaxy cluster, which has resulted in succesful detections of a much fainter galaxy population \citep[e.g.][]{Zheng2014,Ishigaki2015,Livermore2017,Coe2019,Salmon2020}.
Ground-based near-infrared telescopes such as the United Kingdom Infrared Telescope Deep Sky Survey (UKIDSS), the Visible and Infrared Survey Telescope for Astronomy (VISTA), and Subaru, have also played a key role \citep[e.g.][]{Kashikawa2004,McLure2009,Bowler2015}.

The commencement of science operations with the James Webb Space Telescope (\textit{JWST}) will launch a new era of high-$z$ galaxy studies. With highly sensitive imaging capabilities with NIRCam and MIRI from 0.6--5 and 5.6--25.5 microns respectively, \textit{JWST} will be able to detect and characterize high-$z$ galaxies in the rest-frame UV and optical in unprecedented detail, and discover fainter and more distant objects than currently possible.
Its spectroscopic capabilities will allow for secure redshift measurements, and physical insights from key emission lines that cannot be detected from the ground.

One of the key science themes for \textit{JWST} is the early Universe, studying reionization and the first galaxies. Many large surveys are planned for Cycle 1, the first year of \textit{JWST} science operations, which will address this goal.
One of the largest surveys is the \textit{JWST} Advanced Deep Extragalactic Survey (JADES), a collaboration between the NIRCam and NIRSpec galaxy assembly  Guaranteed Time Observations (GTO) teams \citep{Rieke2019,Bunker2021}.
Two key surveys which will be public to the community immediately are the COSMOS-Web Program \citep{Kartaltepe2021} and the Cosmic Evolution Early Release Science Survey \citep[CEERS;][]{Finkelstein2017}.
Surveys such as the Prime Extragalactic Areas for Reionization and Lensing Science (PEARLS) North Ecliptic Pole (NEP) Time-Domain Field \citep{Windhorst2017,Jansen2018} are expected to build up in depth over the lifetime of the mission, due to their importance for studying time-domain phenomena with multiple epochs.
Deep observations of cluster lensing fields is also a promising avenue that will be extensively explored \citep[e.g][]{Willott2017a,Windhorst2017}.
Many programs will also target individual high-$z$ sources that have been discovered by other facilities, to provide detailed new information about their physical properties \citep[e.g][]{Finkelstein2021,Coe2021}.
Overall, this will be a transformative period of new data and discoveries.

Given the importance of these data sets, it is imperative to have a range of theoretical predictions of high-$z$ galaxy properties, to which we can compare the observations. 
Theoretical predictions are also useful to aid in designing the observing strategies used for future observations, both for future cycles with \textit{JWST} but also for planning upcoming mission surveys such as with the \textit{Euclid} and Nancy Grace \textit{Roman} Space Telescopes. 

Previous studies have made estimates for the number of galaxies expected to be found by various \textit{JWST} surveys.
\citet{Yung2018} used the Santa Cruz semi-analytic model to estimate the luminosity function that could be measured by \textit{JWST}, and showed the expected number of galaxies per volume above two detection limits.
\citet{Vogelsberger2020} used the luminosity function from Illustris-TNG, extrapolated to lower magnitudes, to estimate the number of galaxies detected with NIRCam above the JADES-Deep, JADES-Medium and CEERS expected magnitude limits.
\citet{Wilkins2022} used the First Light And Reionisation Epoch Simulations \citep[FLARES;][]{Lovell2020,Vijayan2021}, which contains a large number of high-$z$ galaxies, to predict the number of $z=10$--15 galaxies that will be detected by various \textit{JWST} surveys. They used the FLARES luminosity functions and assumed 100\% completeness down to the 10$\sigma$ point-source depths, and predicted that in \textit{JWST} Cycle 1 approximately 600 galaxies should be identified at $z > 10$.

These studies give useful predictions of the average number of galaxies expected in each survey. However, \citet{Steinhardt2021} considered the effect of cosmic variance on the estimated number counts, using models and observed luminosity functions, and found that this will be the dominant contribution to the uncertainty in high-$z$ galaxy luminosity functions. 
\citet{Behroozi2020} used the empirical model \textsc{UniverseMachine} to produce mock galaxy catalogues and light cones. Based on a simple magnitude cut, they made predictions for $z>10$ galaxy number counts for a range of \textit{JWST} surveys, with a total number of $z>10$ galaxies of 210--972 based on a number of survey realisations. 
However, the expected number counts from all of these studies are based on the luminosity function and a simple 100\% completeness cut at some magnitude limit.

An alternative approach is to create mock images of various \textit{JWST} surveys, to investigate not only the number of sources above some magnitude in a field, but how many may be extracted and correctly characterized.
\citet{Williams2018} created a galaxy catalogue and software for mock image generation based on a phenomenological model, called the JAdes extraGalactic Ultradeep Artificial Realisations (JAGUAR) catalogue. These catalogues take observed UV luminosity and stellar mass functions to create a detailed empirical model of galaxies from $z=0.2$--15.
They made specific simulations of the JADES survey, and predicted that it will discover 1000s of galaxies at $z\geq6$, and 10s of galaxies at $z\geq10$, with $m\lesssim 30$ mag.
By extracting sources from detailed mock images created using the JAGUAR extragalactic catalogues, other studies have explored: the impact of the observing strategy and analysis choices on successfully identifying high-$z$ galaxies from low-z interlopers \citep{Hainline2020}; source blending and how accurately galaxy properties can be recovered
\citep{Kauffmann2020}; and the effectiveness of medium-band filters for improving galaxy property measurements \citep{RobertsBorsani2021}.

\citet{Yang2021} simulated very realistic MIRI images of the CEERS survey, and ran a comprehensive photometric pipeline on their images, including mitigating cosmic rays and a variable background subtraction. They used these to determine how well the MIRI imaging can constrain the properties of high-$z$ galaxies.
\citet{Drakos2022} presented the Deep Realistic Extragalactic Model (DREaM) for creating synthetic galaxy catalogues, which uses an empirical model to create galaxy catalogues for galaxies up to $z\sim12$. These catalogues were used to create detailed lightcones and mock images of a deep survey with \textit{Roman}. \citet{Drakos2022} predicted that a 1 degree$^2$ \textit{Roman} ultra-deep field to $\sim30$ mag could detect more than $10^6$ $M_{UV}<-17$ mag galaxies, with more than $10^4$ at $z > 7$.

These studies provide comprehensive predictions of the number of galaxies various observing strategies will detect, and how well their physical properties can be extracted from the images. However, these simulations all assume that the galaxies are S\'ersic profiles \citep[or point sources for objects unresolved in MIRI in][]{Yang2021}.
Realistically, high-$z$ galaxies are generally clumpy \citep{Jiang2013b,Bowler2016}, and so will not follow a simple S\'ersic profile.
In this work we create mock images of more realistic high-$z$ galaxies from the \BlueTides hydrodynamical simulation. We create a catalogue of images of $\sim100,000$ galaxies in \BlueTides at $z=7$--12, simulating observations with \textit{JWST}, \textit{HST}, \textit{Roman}, \textit{Euclid}, Subaru and VISTA. As opposed to the detailed models above which create full mock image fields, including galaxies at a range of redshifts and even foreground stars, we image only cut-outs of individual galaxies. These images will be useful for studying the properties of high-$z$ galaxies from the \BlueTides simulation, such as their morphologies, as well as making detailed predictions for upcoming surveys. In this work, we focus on predictions of the performance of \textit{JWST}, investigating the fraction of galaxies detected under a range of exposure times with different NIRCam and MIRI filters. We also make predictions for specific \textit{JWST} surveys, such as JADES-Medium, COSMOS-Web, CEERS, PEARLS NEP and PEARLS GOODS-S, predicting the number of expected high-$z$ galaxies and considering the effect of cosmic variance. 

This paper is structured as follows.
In Section \ref{sec:Simulation} we describe our technique for making images with \BlueTidesns, and give an overview of the publicly available mock image catalogue. In Section \ref{Sec:JWSTPredictions} we perform photometry on a range of mock images. We compare various telescopes in Section \ref{Sec:VariousTelescopes}. Focusing on \textit{JWST}, we investigate the completeness of observations with different filters and exposure times (Section \ref{Sec:ExposureTimes}), and make specific predictions for the number of high-$z$ galaxies detected in a range of upcoming \textit{JWST} surveys (Section \ref{Sec:Surveys}). We include a discussion in Section \ref{Sec:Discussion}, and conclude with a summary in Section \ref{Sec:Conclusion}.
The cosmological parameters used throughout are from the nine-year Wilkinson Microwave Anisotropy Probe \citep[WMAP;][]{Hinshaw2013}: $\Omega_M=0.2814$, $\Omega_\Lambda=0.7186$, $\Omega_b=0.0464$, $\sigma_8=0.820$, $\eta_s=0.971$ and $h=0.697$.

\section{BlueTides Mock Images}
\label{sec:Simulation}
Here we give a brief overview of the \BlueTides simulation (Section \ref{sec:BlueTides}), how galaxy spectral energy distributions (SEDs) are modelled (Section \ref{sec:SED}), and the selection of our galaxy sample (Section \ref{Sec:GalaxySample}). We then detail how we created the mock images available in the \BlueTides Mock Image Catalogue in Section \ref{sec:MockImages}.

\subsection{\BlueTides}
\label{sec:BlueTides}
\BlueTides is a large-scale cosmological hydrodynamical simulation, simulating the evolution of galaxies from $z=99$ to $z=7$ \citep{Feng2015,Ni2019}. The extreme volume of \BlueTidesns, $(400/h ~\rm{cMpc})^3$, allows for a statistical study of bright galaxies in the early Universe. The simulation contains $2\times 7040^{3}$ particles, with dark matter, gas, and star particle initial masses of $1.2 \times 10^7/h~ M_{\odot}$, $2.4 \times 10^6/h~ M_{\odot}$, and $6\times10^{5}/h~ M_\odot$ respectively. The gravitational softening length is $\epsilon_{\rm grav} = 1.5/$h ckpc (0.24 kpc at z = 8).

\BlueTides includes a range of prescriptions to model sub-grid physical processes,
for example gas \citep{Katz} and metal line cooling~\citep{Vogelsberger2014},
multi-phase star formation  \citep{SH03,Vogelsberger2013}
including effects from the formation of molecular hydrogen \citep{Krumholtz}, supernova feedback \citep{Okamoto}, reionization \citep{Battaglia}, and black hole growth and active galactic nuclei (AGN) feedback (\citealt{SDH2005,Matteo2005}; see \citealt{DeGraf2012a} and \citealt{DeGraf2015}).
For full details, see the original \BlueTides paper \citet{Feng2015}.

For this work, we extract galaxies for imaging analogously to \citet{Marshall2022}, with the reader referred there for full details.
Galaxy haloes are identified in \BlueTides using a friends-of-friends (FOF) algorithm. To identify individual halo galaxies, we locate each black hole in the halo, and assume each resides at the centre of a galaxy. Black holes are seeded in 
dark matter haloes when they reach a threshold mass of $M_{\rm{Halo,seed}} = 5 \times 10^{10} /h~ M_\odot$, at the location of the densest particle \citep{DeGraf2015}. 

The \BlueTides galaxy properties, such as the star formation density, stellar mass function, and UV luminosity function, have been shown to match current observational constraints at $z=7$, 8, 9 and 10 \citep{Feng2015,Waters2016,Wilkins2017,Marshall2020,Ni2019}.

\subsection{Spectral Energy Distribution Modelling}
\label{sec:SED}
Modelling of the spectral energy distributions (SEDs) of galaxies in \BlueTidesns, including the effects of nebular emission and dust attenuation, is described in detail in \citet{Wilkins2016,Wilkins2017,Wilkins2018,Wilkins2020} and \citet{Marshall2022}. 

Each star particle in a galaxy is assigned an intrinsic stellar SED according to its age and metallicity from the Binary Population and Spectral Synthesis model \citep[BPASS, version 2.2.1;][]{Stanway2018}, assuming a modified Salpeter initial mass function with a high-mass cut-off of $300 M_\odot$. Star particles are then associated with a \ion{H}{II} region (and nebular continuum and line emission) using the \textsc{Cloudy} photo-ionisation code \citep{Ferland2017}, assuming that: the metallicity $Z$ of the \ion{H}{II} region is identical to that of the star particle, the hydrogen density is
100 cm$^{-3}$ \citep{Osterbrock2006}, and no Lyman-continuum photons escape.

We consider dust attenuation from the interstellar medium (ISM). For each star particle we calculate the line-of-sight density of metals $\rho_{\rm metal}$ and convert this to a dust optical depth:
\begin{equation}
 \tau_{\rm ISM}= \kappa \left(\frac{\lambda}{5500\text{\normalfont\AA}}\right)^{\gamma}  \int_{z'=0}^{z} \rho_{\rm metal}(x,y,z')dz' 
\end{equation}
where we assume that $\gamma=-1$, i.e.  $\tau_\lambda \propto \lambda^{-1}$, and use $\kappa=10^{4.6}$, which is calibrated against the observed galaxy UV luminosity function at redshift $z=7$ \citep{Marshall2020,Ni2019}.
For star particles with ages less than 10 Myr, we also assume dust contribution from a birth cloud, with optical depth
$\tau_{\rm BC}= 2 \left(Z/Z_\odot\right)
\left(\lambda/5500\text{\normalfont\AA}\right)^{\gamma}$,
where we assume $\gamma=-1$.

This model assumes a constant dust-to-metal ratio with redshift.
We do not model a varying $\kappa$ with redshift, as there is significant uncertainty in the $z\geq7$ galaxy UV luminosity functions used for the calibration. By calibrating $\kappa$ at $z=7$, we find good agreement with the observed galaxy UV luminosity functions at $z=7$--10 (see Figure \ref{LFs}), suggesting that a non-evolving dust-to-metal ratio is a reasonable assumption for this work.
However, if the dust is (under) overestimated at higher redshifts, we would expect our predictions to (over) underestimate the number of high-$z$ galaxies that could be detected.
 
Note that throughout this work we consider only the stellar emission, and do not include the emission from AGN.
The core of high-$z$ galaxies can be very dusty, strongly attenuating the AGN \citep[e.g.][]{Bowler2021}, hence justifying this omission \citep[see e.g.][]{Roper2022}.
We have also studied the host galaxies of \BlueTides quasars specifically by creating mock \textit{JWST} images in \citet{MarshallBTpsfMC}.

\subsection{Galaxy Sample}
\label{Sec:GalaxySample}

\begin{figure*}
\begin{center}
\includegraphics[scale=1]{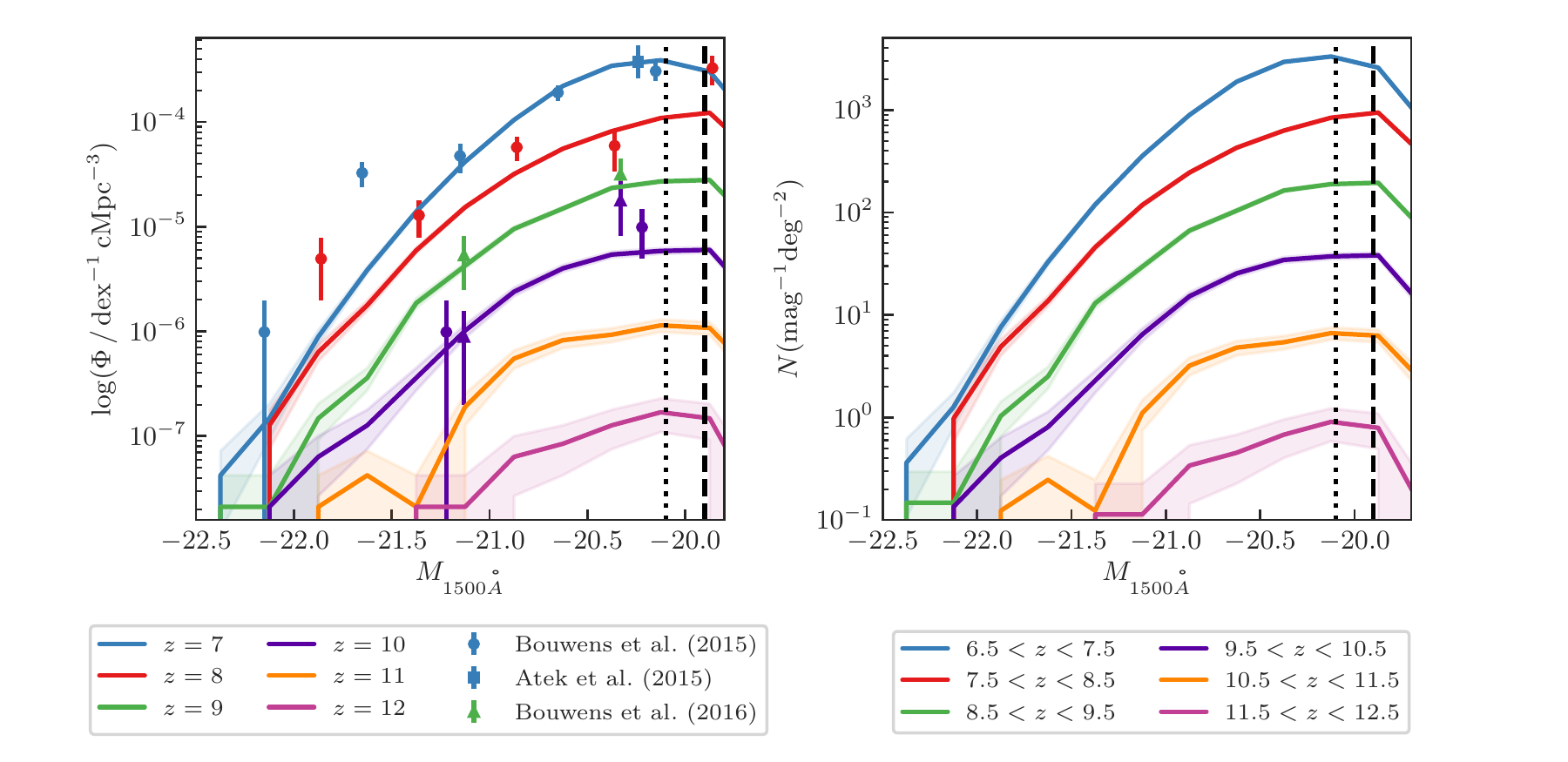}
\vspace{-0.5cm}
\caption{Left: The UV luminosity function for the \BlueTides galaxy sample at $z=7$, 8, 9, 10, 11 and 12 (coloured lines), compared to observations from \citet{Bouwens2015}, \citet{Atek2015} and \citet{Bouwens2016}. The symbols for each set of observations are shown in the legend, and they are coloured according to their redshift, matching the \BlueTides colours (see legend).
Right: The number of galaxies in \BlueTides per magnitude per square degree field of view. This is given over an integer redshift range, converted from the specific redshift snapshots via Equation \ref{Eqn:Conversion}. The completeness limit of \BlueTidesns, due to its resolution, is marked in dotted black lines for $z=7$ ($M_{\textrm{1500\AA}}=-20.1$ mag) and dashed black lines for $z>7$ ($M_{\textrm{1500\AA}}=-19.9$ mag).}
\label{LFs}
\end{center}
\end{figure*}

In this work we consider a luminosity-limited galaxy sample, selecting galaxies with dust-attenuated far-UV (FUV) luminosity $L_{\textrm{1500\AA}}>10^{28.5} {\textrm{erg/s/Hz}}$ or $M_{\textrm{1500\AA}}<-19.65$ mag, as
\BlueTides is incomplete for lower luminosity galaxies due to its resolution.
This sample contains
31 galaxies at $z=12$, 244 galaxies at $z=11$, 1,279 at $z=10$, 5,606 at $z=9$, 22,144 at $z=8$ and 71,052 at $z=7$.
The UV luminosity function for these galaxies is shown in Figure \ref{LFs}, alongside the observations of \citet{Bouwens2015}, \citet{Atek2015} and \citet{Bouwens2016}. 
At $z=7$ \BlueTides is complete to $M_{\textrm{1500\AA}}=-20.1$ mag, and at $z>7$ it is complete to $M_{\textrm{1500\AA}}=-19.9$ mag, with the luminosity function turning over and decreasing for fainter magnitudes (see Figure \ref{LFs}). Although we do not have a higher resolution \BlueTides simulation with which to compare, this turn-over is a clear indicator of where the simulation would no longer converge due to numerical resolution. Thus, throughout this work we do not consider galaxies below this convergence/completeness limit, to ensure the accuracy of our results.

In Figure \ref{LFs} we also show the total number density per square degree of \BlueTides galaxies in a redshift interval $z'-0.5<z<z'+0.5$. 
This is calculated by assuming that
\begin{align}
\label{Eqn:Conversion}
N (z'-0.5<z<z'+0.5) = N(z=z') \times \frac{D_C(z'-0.5\textrm{ to }z'+0.5)}{400/h}
\end{align}
where $N (z'-0.5<z<z'+0.5)$ and $N(z=z')$ are the number of galaxies in the depth $z'-0.5<z<z'+0.5$ and in the simulation snapshot at $z=z'$, respectively. The co-moving radial distance $D_C$ between $z=6.5$ and 7.5 is 358.3 cMpc, $z=7.5$--8.5 is 300.4 cMpc, $z=8.5$--9.5 is 256.5 cMpc, $z=9.5$--10.5 is 222.32 cMpc, $z=10.5$--11.5 is 195.1 cMpc, and $z=11.5$--12.5 is 173.0 cMpc.  
The \BlueTides box has a depth of $400/h=573.9$ cMpc, deeper than one integer redshift window at these high redshifts, and so this conversion results in a reduction of the number of galaxies.
This is a coarse estimate that assumes galaxies do not evolve in $z'-0.5<z<z'+0.5$, and can be adequately approximated in this interval by the population at $z=z'$. This assumption is likely accurate at the highest redshifts where the time interval is small, although by $z=7$ this may not be the case.

\begin{table*}
\vspace{-0.5cm}
\caption{Table of the simulated telescope filters, specifying their pixel scale and field of view (FOV). The pixel scale has a sub-sampling of the native pixel scale by a factor of 2. \textit{Euclid}, VISTA, and Subaru have larger FOVs due to their low resolutions.}
\begin{tabular}{llllc}
\hline 
Telescope & Instrument & Filters & Pixel Scale ('') & Image FOV (pkpc$^2$) \\ 
\hline 
\textit{JWST} & NIRCam (Short Wavelength) & F090W, F115W, F150W, F200W & 0.0155 & $6 \times 6$ \\ 
 & NIRCam (Long Wavelength) & F277W, F356W, F410M, F444W & 0.0315 & $6 \times 6$ \\ 
& MIRI & F560W, F770W & 0.055 & $6 \times 6$ \\ 
\textit{HST} & WFC3 & F105W, F125W, F140W, F160W & 0.065 & $6 \times 6$ \\ 
\textit{Roman} & WFI & F087, F106, F129, F146, F158, F184 & 0.055 & $6 \times 6$ \\ 
\textit{Euclid} & NISP & $Y$, $J$, $H$ & 0.15 & $10 \times 10$ \\ 
VISTA & VIRCam & $Z$, $Y$, $J$, $H$, $Ks$ & 0.17 & $10 \times 10$ \\ 
Subaru & HSC & $z$, $y$ & 0.085 & $10 \times 10$ \\ 
\textit{Spitzer}* & IRAC & Ch1, Ch2 & - & - \\ 
\hline 
\end{tabular} 
\label{table:Filters}
\begin{flushleft}
\footnotesize{$^*$ 
Note that we provide only the \textit{Spitzer} fluxes and not the full set of images, as galaxies are unresolved.}
\end{flushleft}
\end{table*}

\subsection{Mock Images}
\label{sec:MockImages}

\begin{figure*}
\begin{center}
\includegraphics[scale=0.7]{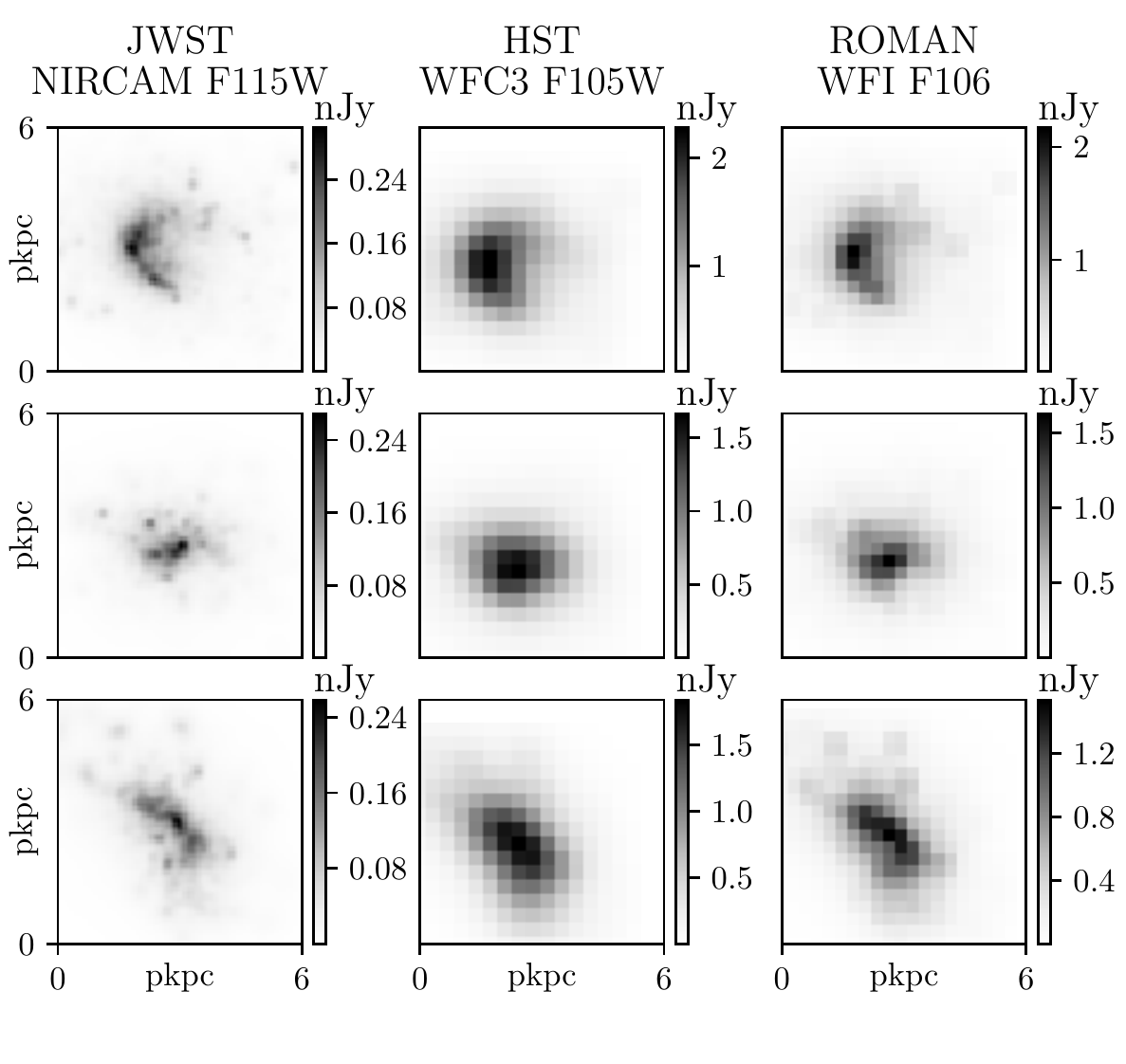}
\hspace{0.4cm}
\includegraphics[scale=0.7]{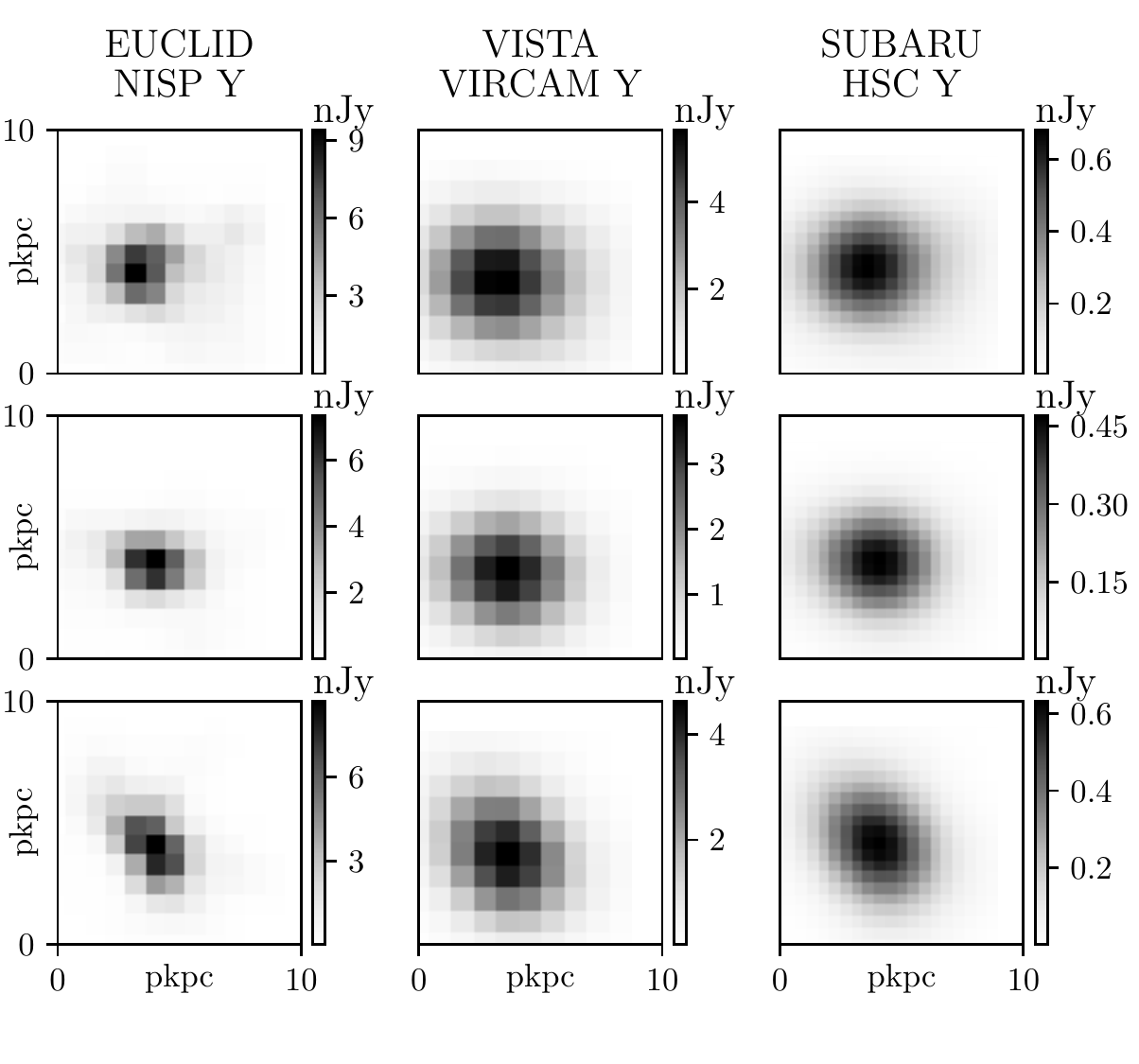}
\vspace{-0.5cm}
\caption{Example $Y$-band images of three $z=7$ \BlueTides galaxies in each of the telescopes. Note that the \textit{Euclid}, VISTA, and Subaru images cover larger area ($10\times10$ pkpc vs $6\times6$ pkpc). These images do \emph{not} contain noise.}
\label{YbandImages}
\end{center}
\end{figure*}

\begin{figure*}
\begin{center}
\vspace{-0.5cm}
\includegraphics[scale=1]{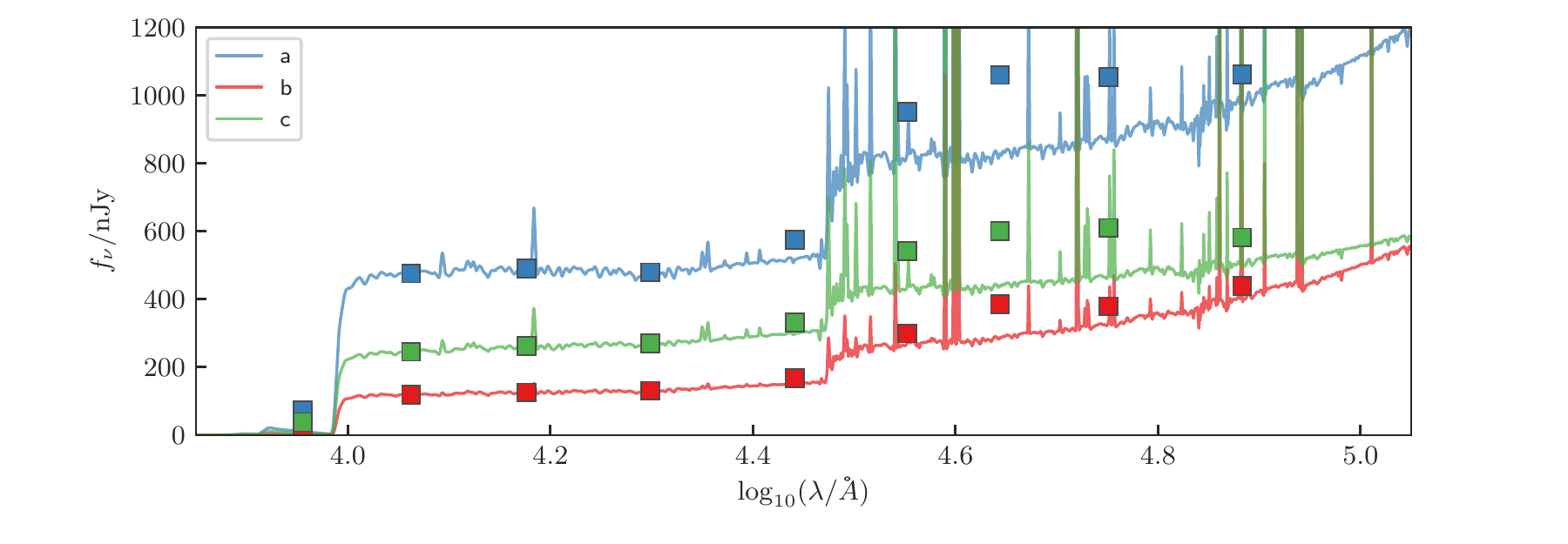}

\hspace*{-0.5cm}\includegraphics[scale=0.56]{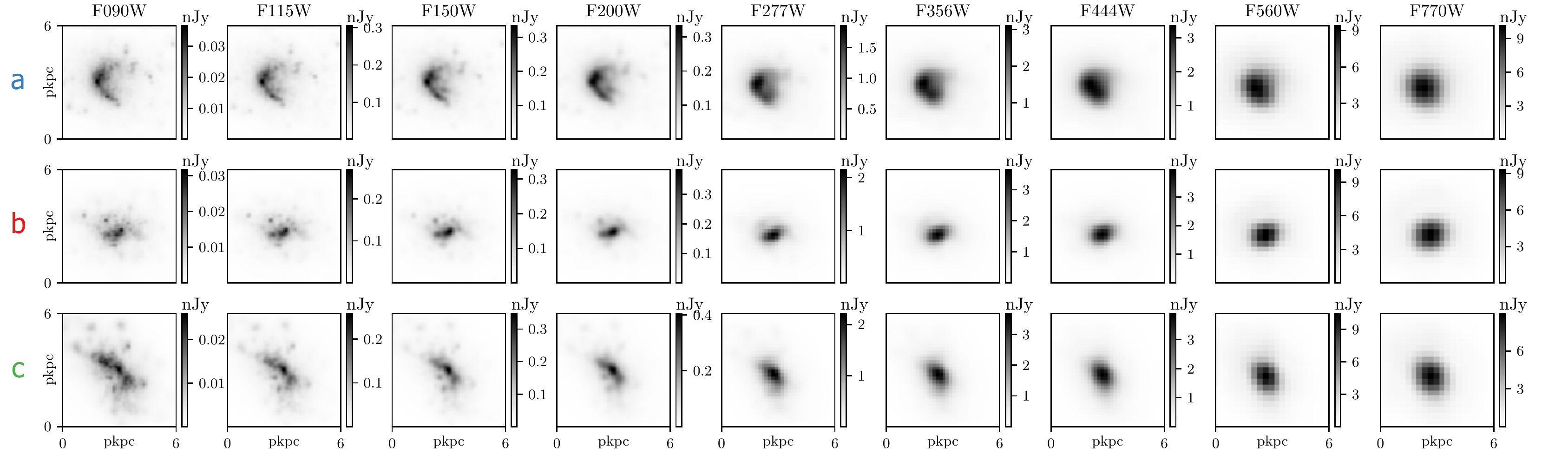}
\caption{Top: Example spectra for three $z=7$ galaxies, alongside their photometry in the seven simulated \textit{JWST} NIRCam filters and two MIRI filters. Bottom: The corresponding mock \textit{JWST} images, not including any noise.}
\label{Plot:ExampleImages}
\end{center}
\end{figure*}

In \citet{Marshall2022} we created 
rest-frame images of the $z\gtrsim7$ \BlueTides galaxies in standard top-hat filters: FUV (1500\AA), 2500\AA, $U$, $B$, $V$, $I$, $Z$, $Y$, $J$ and $H$, \textit{not} including instrumental effects such as a point spread function (PSF) or noise.
In \citet{MarshallBTpsfMC} we created mock \textit{HST} and \textit{JWST} images, including the PSF, pixel scale, and estimated noise, however we considered only the galaxies which host quasars at $z=7$. 

Here we extend these previous studies and create mock images of a large sample of $\sim100,000$ \BlueTides galaxies from $z=7$--12 for a range of current and upcoming telescopes. A catalogue of these mock images is publicly available via the Mikulski Archive for Space Telescopes\footnote{ \href{https://doi.org/10.17909/ER09-4527}{doi:10.17909/ER09-4527}}.
These images consider the instrument PSF and pixel scale, however they are produced without noise so that they can be adapted for each specific use case. Noise (both sky and shot noise) is able to be simply added in post-processing by code which accompanies the catalogue\footnote{\href{https://github.com/madelinemarshall/BlueTidesMockImageCatalogue}{github.com/madelinemarshall/BlueTidesMockImageCatalogue}}; we investigate adding noise in Section \ref{Sec:JWSTPredictions}. 

For this work we simulate \textit{JWST}, \textit{HST}, \textit{Euclid}, \textit{Roman}, VISTA, and Subaru images, with a full list of the instruments and filters given in Table \ref{table:Filters} and visualized in Figure \ref{Filters}. We generally consider the filters for each instrument that are red-ward of the Lyman-limit at $z=7$ at 0.73 microns. 
Filters blue-ward of Lyman-alpha at $z=7$ at 0.97 microns, for example the $z$-band, contain very minimal flux, although they are included as they may be of interest for comparisons. We generally consider only the wide-band filters, which are key for finding faint objects; for \textit{JWST} NIRCam we also simulate F410M, the widest of the medium-band filters, which is popular for high-$z$ galaxy surveys due to its complementarity to F444W.

As in \citet{Marshall2022} our images are of the $6\times6$ kpc field of view (FOV) around each galaxy, except for \textit{Euclid}, VISTA, and Subaru, which have much wider PSFs and so require a larger FOV of $10\times10$ kpc to contain the galaxy flux.
We bin the image to a pixel scale of 0.5 times the native pixel scale. This assumes that given sufficient dithering in the observations, the final image can sub-sample the original pixel scale by a factor of 2. This is a useful strategy particularly for \textit{JWST} NIRCam which will under-sample the PSF at some wavelengths. The catalogue images can be re-sampled to a larger pixel scale if required.
These image properties are listed in Table \ref{table:Filters}.

The SED from each star particle in the FOV is convolved with the instrument filter transmission curve (Figure \ref{Filters}), to determine its flux in the filter. We smooth the light from each star particle adaptively with the smoothing scale (full width at half maximum, FWHM, of the Gaussian) equal to the distance to the 8th nearest neighbour (in 3D). These smoothed fluxes from each star particle within the galaxy are combined to determine the flux in each pixel.

Finally, the resulting image is convolved with the PSF of the instrument. The PSFs are obtained via TinyTim for \textit{HST} \citep{Krist_2011}, and WebbPSF for \textit{JWST} and \textit{Roman}  \citep{Perrin2015}. \textit{Euclid} NISP is assumed to have a Gaussian PSF with FWHM of $0\farcs175$ in the $Y$-band, $0\farcs24$ in $J$, and $0\farcs28$ in $H$ \citep{EuclidPSF}.
Ground-based Subaru and VISTA are assumed to have Gaussian PSFs with FWHM of $0\farcs60$ and $0\farcs66$ respectively, corresponding to the typical site seeing \citep{SubaruSeeing,VISTAseeing}.
We also include in the catalogue separate images that have not been convolved with the PSF, which could be convolved with the true PSF measured for the instrument as opposed to the models, or used in image simulation software that requires an un-convolved image such as \textsc{MIRAGE} for \textit{JWST} mock observations \citep{MIRAGE2022}, for example. Throughout this work, however, we only use the PSF-convolved images.

The effective spatial resolution of \BlueTides is the gravitational softening length  $\epsilon_{\textrm{grav}} = 1.5/h$ ckpc, which corresponds to 0.269 pkpc, or $0\farcs05$ at $z=7$ and $0\farcs07$ at $z=12$ in this cosmology. The resolution of \BlueTides is thus well-matched to \textit{JWST}, which has a resolution of $0\farcs05$ at $1.5\mu$m, with all other instruments considered here having much lower resolution.

The images of the galaxies are made in the `face-on' direction, as determined by the angular momentum of particles in the galaxy \citep[see][]{Marshall2020}.
The quoted luminosity of a galaxy is the total luminosity in the corresponding image.
Output images are in \textsc{fits} file type, in units of nJy.

In Figure \ref{YbandImages} we show example images of three $z=7$ \BlueTides galaxies in each telescope, in the Y-band or closest equivalent. We also show example mock \textit{JWST} NIRCam and MIRI images of these three $z=7$ galaxies, alongside their spectra in Figure \ref{Plot:ExampleImages}.

\section{Predictions for High-z Galaxy Observations}
\label{Sec:JWSTPredictions} 
In this Section we consider a range of observing strategies, and investigate which of the \BlueTides galaxies can be successfully detected in the corresponding mock images. To do this, we add background noise expected from different exposure times and specific upcoming surveys, and perform photometric source extraction on the galaxy images, as described in Section \ref{sec:Photometry}. 

We first compare the success of each of the telescopes for observing the same $z=7$ galaxy sample with the same exposure time (Section \ref{Sec:VariousTelescopes}).
We then make predictions for \textit{JWST}, considering a range of filters and exposure times in Section \ref{Sec:ExposureTimes}. Finally, we make detailed predictions for detections of high-$z$ galaxies in planned \textit{JWST} surveys in Section \ref{Sec:Surveys}.

\subsection{Photometric Source Extraction}
\label{sec:Photometry}
To create our mock telescope images, we add noise based on the expected depth for each exposure time.
The noise $\sigma$ for the mock \textit{JWST} images is estimated from the predicted $10\sigma$ sensitivity of \textit{JWST} for the corresponding exposure time, from the values provided by \citet{NIRCamImaging,MIRI2017}. These use a circular photometric aperture 2.58 pixels in radius. These noise estimates assumed 1.2 times the minimum zodiacal light background at RA = 17:26:44, Dec = -73:19:56 on June 19, 2019 \citep{NIRCamImaging}.
For the other telescopes we use their available Exposure Time Calculators (ETCs) to determine $\sigma$ from the 5$\sigma$ or 10$\sigma$ sensitivities at the given exposure time. For each calculation we consider a point source with a flat spectrum in $F_\nu$.
For \textit{HST} \citep{WFC3ETC} we assume an aperture of radius $0\farcs2$, and 4 dither positions.
For Subaru \citep{HSCETC} and VISTA \citep{VISTAETC} we assume an aperture radius of 2$''$. As for the PSFs, we assume a seeing of $0\farcs6$ and $0\farcs66$, respectively.
For \textit{Roman} \citep{Pickering2016}  we assume an aperture of radius $0\farcs286$, and a background annuli of $0\farcs4$--$0\farcs5$. The readout settings we chose for the 10ks exposure are `deep8' with 20 groups and 7 integrations.
All other settings in the ETCs are set to their default values.
We assume that this noise follows a Gaussian distribution. We also include shot noise from the source flux, following a Poisson distribution.

We use \textsc{Photutils} \citep{photutils} to extract the galaxy from each cut-out image, following the procedure outlined in \citet{Photometry2022}, particularly the \textsc{run_detect_sources} function. For a detection, we require 5 connected pixels at/above the threshold level of 2$\sigma$ times the background RMS. 

NIRCam's short and long wavelength filters, and MIRI, have different native pixel scales of $0\farcs031$, $0\farcs063$ and $0\farcs11$ respectively. Our images are sub-sampled to a factor of 2, or pixel scales of $0\farcs0155$, $0\farcs0315$ and $0\farcs055$ respectively (Table \ref{table:Filters}). We find that a re-sampled image with larger pixel scale aids in detection of fainter objects for the shorter wavelength filters. We re-bin the NIRCam detection images by a factor of 4 and 2 to be on a scale of 
$0\farcs062$ and $0\farcs063$ for the short- and long-wavelength filters respectively, so that these and the MIRI images are on a similar pixel scale.

For the other telescopes, we resample by a factor of 2, reverting to the native pixel scale.

We note that we use only this standardised \textsc{Photutils} galaxy extraction algorithm throughout this work. High-$z$ galaxy searches may likely consider alternative or additional source detection algorithms, such as \textsc{Source Extractor} \citep{Bertin1996} or \textsc{ProFound} \citep{Robotham2018}, for example. One could also carefully optimise the parameters in these algorithms to improve the detection of galaxies in specific sets of images. The choice of pixel scale and wavelengths used for the detection images are expected to have an effect on the success of the strategy, and image stacking could also be used to improve the detectability of fainter galaxies.
In addition, with more modestly sized galaxy samples from true images, each galaxy candidate could be reasonably visually confirmed, 
which is not possible for our sample of $\sim100,000$ high-$z$ galaxies, which may alter the detection strategy. Our approach also does not consider the difficulty of determining the photometric redshifts of any detected galaxies, which would be necessary for real images.
Overall, careful and optimised galaxy extraction techniques may be more successful than expected by the more simplistic, standard predictions here, nonetheless our approach offers useful insights and comparisons.

\subsection{Comparison between Telescopes}
\label{Sec:VariousTelescopes}

In this Section we compare each of the telescopes for which we have created mock images in the catalogue. We add noise corresponding to exposure times of 10,000s, in the $Y$-band or closest equivalent filter: F115W for \textit{JWST}, F105W for \textit{HST}, and F106 for \textit{Roman}. 
We note that \textit{Euclid} does not have a General Observer program like the other instruments, instead having a fixed survey strategy with specific exposure depths, and so we do not include \textit{Euclid} in this comparison.

\begin{figure}
\begin{center}
\includegraphics[scale=0.9]{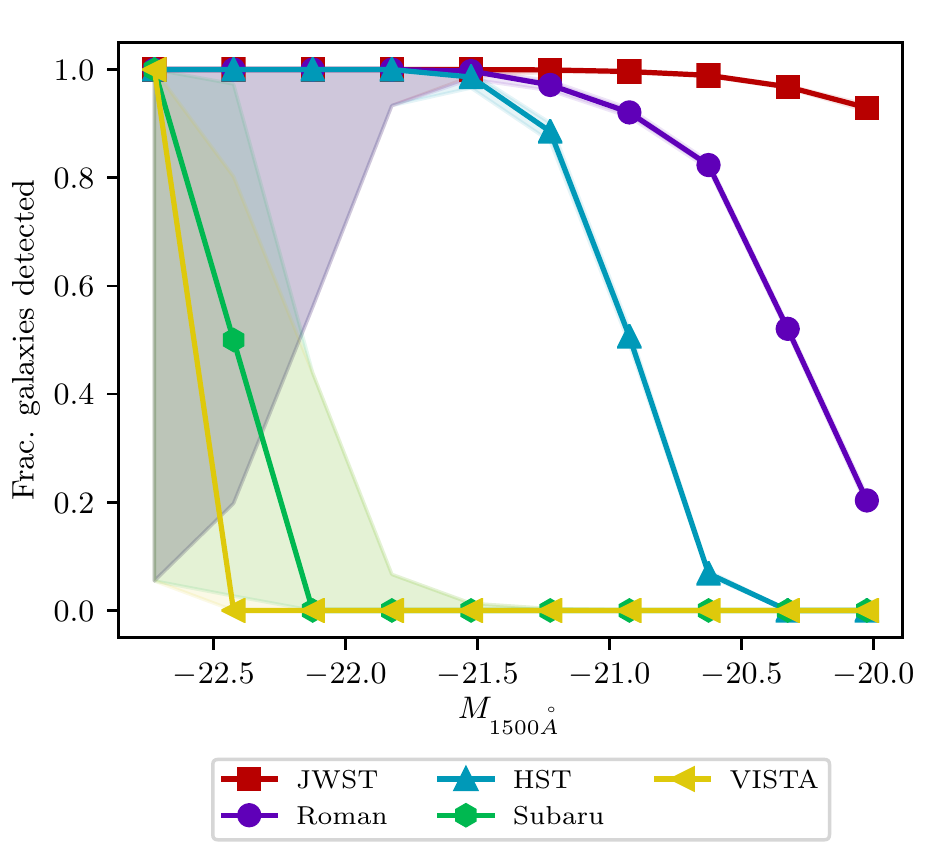}
\caption{The fraction of $z=7$ galaxies detected as a function of their rest-frame UV magnitude, for various telescopes. These assume an exposure time of 10ks, in the $Y$-band or closest equivalent filter: F115W for \textit{JWST}, F105W for \textit{HST}, and F106 for \textit{Roman}. Bins have a width of 0.3 mag. Error bars are 95\% Binomial confidence intervals calculated using the Wilson score interval. Only one galaxy is contained in the brightest magnitude bin, resulting in large associated uncertainty on the true detectable fraction.}
\label{DetectionFraction_telescopes}
\end{center}
\end{figure}

In Figure \ref{DetectionFraction_telescopes} we show the fraction of $z=7$ galaxies detected in these 10ks $Y$-band images, for each telescope, as a function of their rest-frame UV magnitude. Note that we consider only galaxies with $M_{\textrm{1500\AA}}<-20.1$ mag at $z=7$ and $M_{\textrm{1500\AA}}<-19.9$ mag at $z>7$, the completeness limit of \BlueTidesns.
The ground-based telescopes Subaru and VISTA have the brightest completeness limits, becoming 95\% complete at $M_{\textrm{1500\AA}}<-22.7$ mag. \textit{HST} F105W reaches at least 1.5 magnitude deeper, becoming 95\% complete at $M_{\textrm{1500\AA}}=-21.3$ mag. \textit{Roman} F106 has a slightly fainter completeness limit than \textit{HST} of $M_{\textrm{1500\AA}}=-21.1$ mag. The \textit{Roman} detection fraction drops more slowly than \textit{HST}, with 50\% detection success at $M_{\textrm{1500\AA}}=-20.3$ mag, 0.6 magnitudes fainter than \textit{HST}.
Finally, \textit{JWST} will have larger detection fractions than each of these telescopes, becoming 95\% complete at $M_{\textrm{1500\AA}}=-20.2$ mag, at least 2.5 magnitudes fainter than VISTA and Subaru, 1.1 magnitudes fainter than \textit{HST} and 0.9 magnitudes fainter than \textit{Roman}. This demonstrates the unprecedented power \textit{JWST} will have in detecting high-$z$ galaxies.

\subsection{\textit{JWST} Predictions for Various Filters and Exposure Times}
\label{Sec:ExposureTimes}

\begin{figure*}
\begin{center}
\includegraphics[scale=0.9]{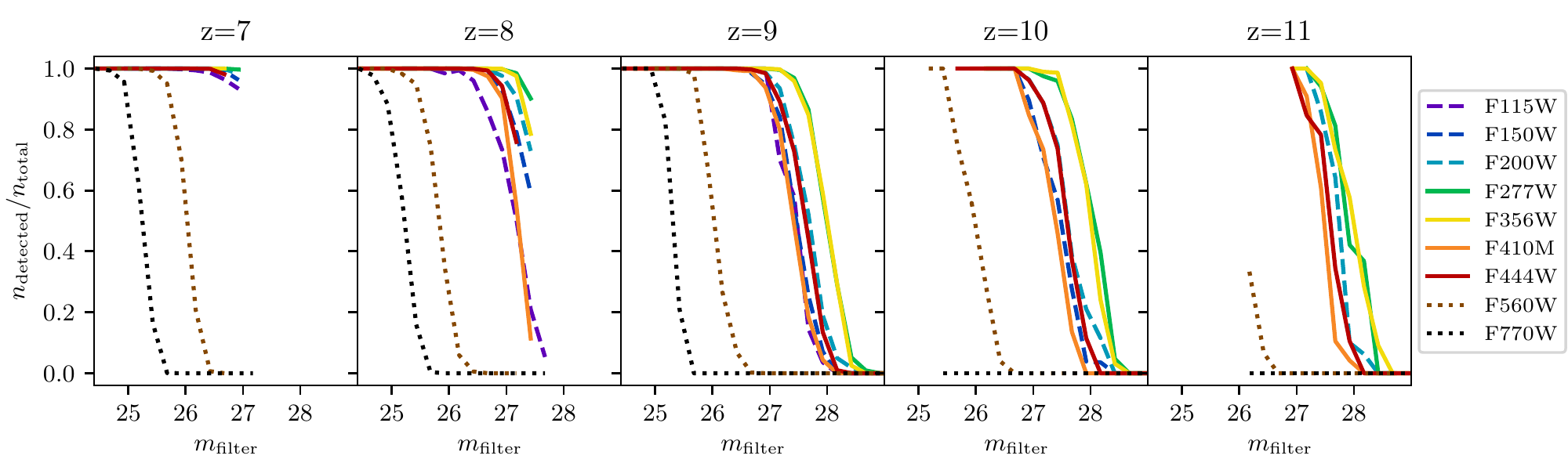}
\vspace{-0.5cm}
\caption{The fraction of galaxies detected as a function of their magnitude in each JWST NIRCam filter, for an exposure time of 10ks, at $z=7$-11. Bins have a width of 0.25 mag. Only bins containing 5 galaxies or more are shown. We show only galaxies above the \BlueTides completeness limit, with $M_{\textrm{1500\AA}}<-20.1$ mag at $z=7$ and $M_{\textrm{1500\AA}}<-19.9$ mag at $z>7$.}
\label{DetectionFraction}
\end{center}
\end{figure*}

\begin{figure*}
\begin{center}
\includegraphics[scale=0.9]{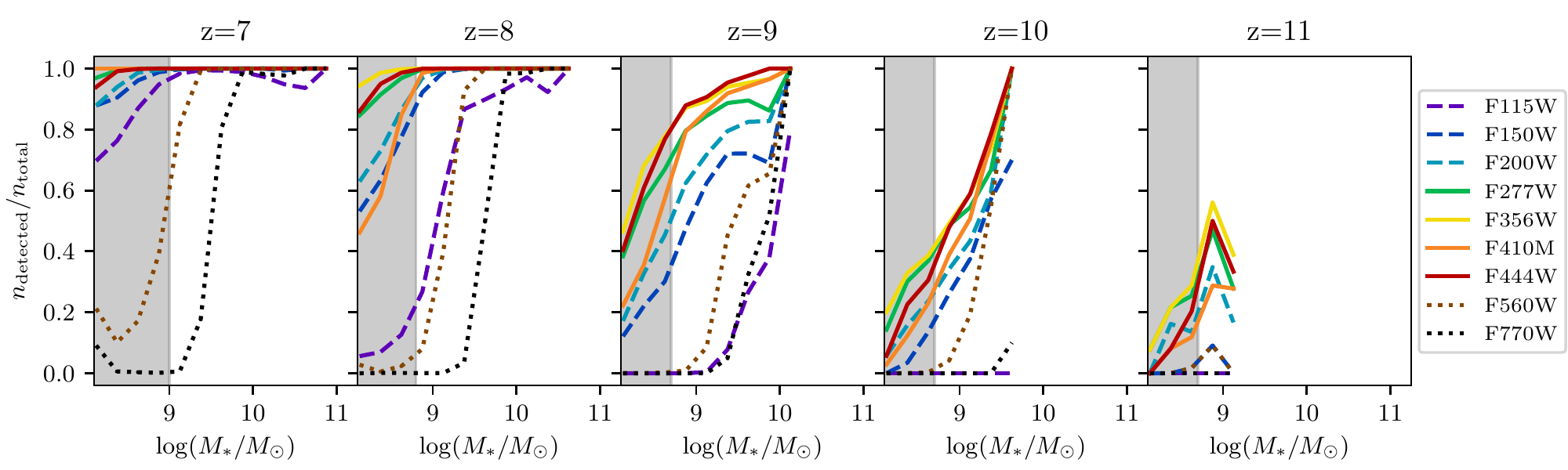}
\vspace{-0.5cm}
\caption{The fraction of galaxies detected as a function of their stellar mass, for an exposure time of 10ks, at $z=7$-11. The regions where the \BlueTides stellar mass function is incomplete are shaded in grey.  Bins have a width of 0.25 mag. Only bins containing 5 galaxies or more are shown. We show only galaxies above the \BlueTides completeness limit, with $M_{\textrm{1500\AA}}<-20.1$ mag at $z=7$ and $M_{\textrm{1500\AA}}<-19.9$ mag at $z>7$.}
\label{DetectionFractionStellarMass}
\end{center}
\end{figure*}

\begin{figure*}
\begin{center}
\includegraphics[scale=0.9]{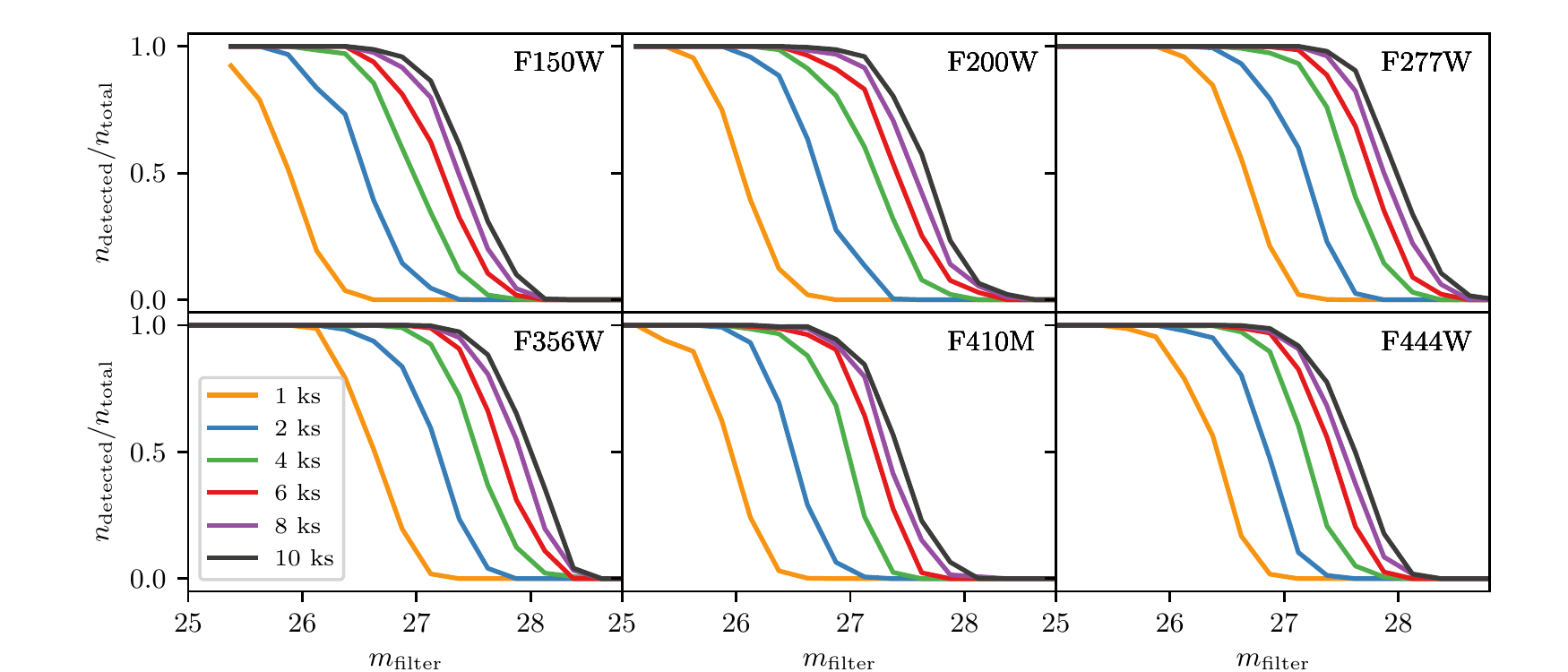}
\caption{The fraction of $z=9$ galaxies detected as a function of their magnitude in each filter, for a range of exposure times (see legend). Bins have a width of 0.25 mag. Only bins containing 5 galaxies or more are shown.}
\label{DetectionFractionExposureTime}
\end{center}
\end{figure*}

To further investigate the expected performance of \textit{JWST}, we create mock images with a range of exposure times: 1, 2, 4, 6, 8, and 10ks, for the full range of simulated filters.
We use these standard exposure times, across the full \BlueTides area, to investigate how the completeness for observations varies with exposure time and filter choice.

In Figure \ref{DetectionFraction} we show the fraction of galaxies at $z=7$--11 that are detected in 10ks images, for the various NIRCam and MIRI filters. We consider only galaxies above the completeness limit, with $M_{\textrm{1500\AA}}<-20.1$ mag at $z=7$ and $M_{\textrm{1500\AA}}<-19.9$ mag at $z>7$.

The average magnitudes at which these 10ks observations become 95\% complete are 26.7, 26.9, 27.1, 27.4, 27.4, 26.9, 27.0, 25.7, and 24.9 mag for F115W, F150W, F200W, F277W, F356W, F410M, F444W, F560W, and F770W, respectively, with a variation of $\pm0.1$ mag between the various redshifts for all filters, except for F115W with $\pm0.2$ mag and $\pm0.3$ mag for F560W.
The predicted 10$\sigma$ point source sensitivities are 
28.6, 28.8, 29.0, 28.5, 28.7, 28.0, 27.9, 26.1 and 25.4 mag, respectively \citep{NIRCam2017,MIRI2017}, 1.9 to 0.5 magnitudes fainter than our 95\% completeness limits. This a reasonable difference given our galaxies are extended; we have verified this difference through simple testing using the \textit{JWST} Exposure Time Calculator\footnote{\url{https://jwst.etc.stsci.edu/}}.
The F277W and F356W filters can detect the faintest galaxies in a fixed exposure time, while the MIRI filters are the least sensitive, becoming incomplete at the brightest magnitudes.

In Figure \ref{DetectionFractionStellarMass} we show the detection fractions as a function of stellar mass, for $z=7$--11. These 10ks NIRCam exposures can detect $>95\%$ galaxies down to $\sim10^9M_\odot$ at $z=7$; the MIRI exposures become incomplete at larger stellar masses, with the F770W filter 95\% complete only to $10^{9.7}M_\odot$. The mass completeness limit increases at $z>7$, with galaxies with given mass becoming more difficult to detect at higher-$z$. Note also that the galaxies redshift out of the F115W band, with Lyman-alpha red-ward of the filter at $z\geq10$.

In Figure \ref{DetectionFractionExposureTime} we show the fraction of $z=9$ galaxies that are detected in images with exposure times ranging from 1--10ks, for each of the NIRCam filters. We chose $z=9$ as at this redshift the galaxies span the widest range in relevant apparent magnitudes, showing the 
completeness fraction from $m_{\textrm{filter}}\simeq25$--29.
Increasing the exposure time from 1 to 10ks is expected to improve the detection depth by approximately 1.25--1.6 magnitudes, with a larger increase for the short-wavelength filters.

\subsection{Predictions for Planned \textit{JWST} High-z Galaxy Surveys}
\label{Sec:Surveys}
In this Section we predict the number of high-$z$ galaxies that will be detected by a range of \textit{JWST} Cycle 1 surveys.

\subsubsection{The \textit{JWST} Surveys Considered}

\begin{table*}
\caption{The key observing information for each of the \textit{JWST} Cycle 1 surveys considered in this work.}
\begin{tabular}{llll}
\hline 
\thead{Survey\\\quad} & \thead{Survey Area\\(square arcmin)} & \thead{NIRCam Filters\\\quad} & \thead{Filter Exposure Times \\(ks)} \\
\hline 
PEARLS GOODS-S & 9.7 &
F090W, F115W, F150W, F200W, F277W, F356W, F410M, F444W
 & 
4.0, 3.8, 2.5, 2.5, 2.5, 2.5, 3.8, 3.8
\\
PEARLS NEP & 64.2 &
F090W, F115W, F150W, F200W, F277W, F356W, F410M, F444W
 & 
2.9, 2.9, 3.35, 3.35, 3.35, 3.35, 2.9, 2.9
\\
CEERS$^*$ & 96.8 &
F090W, F115W, F150W, F200W, F277W, F356W, F444W
 & 
All 2.9 \\
COSMOS-Web & 2160.0 &
F115W, F150W, F277W, F444W
 & 
All 0.774 \\
JADES-Medium & 190.0 &
F090W, F115W, F150W, F200W, F277W, F356W, F410M, F444W
 & 
All 12.0 \\
\hline
\end{tabular} 
\label{Table:Surveys}
\begin{flushleft}
\footnotesize{$^*$ Note that the CEERS survey contains 10 NIRCam pointings of varying depth, however we simulate the minimum exposure time that will cover the full survey area.}
\end{flushleft}
\end{table*}

In this work we consider five extragalactic surveys that will be conducted in \textit{JWST} Cycle 1. These surveys will implement varying strategies to study high-$z$ galaxies, with a range of survey areas and exposure times. We list the area and exposure times we assume for each survey in Table \ref{Table:Surveys}.

JADES \citep{Rieke2019} is an ambitious imaging and spectroscopic deep-field survey which aims to study the formation and evolution of galaxies from $z\sim2$ to $z\geq12$ using a combination of 950 hours of NIRCam, NIRSpec and MIRI data. 
The NIRCam imaging component will survey two fields, a `Medium' field of area 190 arcmin$^2$ to a $10\sigma$ point source limit of 28.8 mag, in both GOODS-S and GOODS-N, and a `Deep' survey which will cover a smaller area of 46 arcmin$^2$ to a $10\sigma$ point source limit of 29.8 mag, centred on the HUDF/GOODS-S. 
In this work we only consider the JADES-Medium survey, as we find that even with its shallower depth it is capable of detecting $\sim100\%$ of \BlueTides galaxies at $z<8.5$ in our magnitude-limited sample of $M_{\textrm{1500\AA}}\lesssim-20$ mag.

CEERS \citep{Finkelstein2017} will observe a field of 100 arcmin$^2$ in the Extended Groth Strip (EGS) with NIRCam, MIRI, and NIRSpec imaging and spectroscopy, aiming to demonstrate successful survey strategies with \textit{JWST}. The NIRCam component is designed to detect a large sample of $z\simeq9$--13 galaxies. We note that for CEERS, the depth of each of the 10 NIRCam pointings is not constant, and so we assume the shallowest depth that covers the full 100 arcmin$^2$ area. 

The COSMOS-Web Survey \citep{Kartaltepe2021} is a large-area survey of 0.6 deg$^2$, designed to study bright galaxies in the early Universe and provide a primary extragalactic legacy dataset. This survey has a much shallower depth, but aims to detect `an order of
magnitude more early Universe galaxies than all other Hubble$+$Webb surveys combined' \citep{Kartaltepe2021}, due to its very large survey area.

The PEARLS program (formerly Webb Medium-Deep Fields) contains several surveys of blank fields \citep{Windhorst2017}. Here we consider the PEARLS NEP field \citep{Jansen2018} and the PEARLS GOOD-S survey.
The NEP field is within \textit{JWST}’s northern continuous viewing zone. The PEARLS NEP program is designed for multi-epoch time-domain observations, and will cover the field with both NIRCam imaging and NIRISS spectroscopy. The imaging is of similar depth to the CEERS field, with 2/3 of the area. The PEARLS GOODS-S survey is a much more modest program of only 5.5 hours, and so may be a more representative example of general Cycle 1 programs.

We note that the \BlueTides simulation is incomplete at faint magnitudes $M_{\textrm{1500\AA}}\gtrsim-20$ mag, due to the resolution of the simulation. As shown below, these medium-depth surveys have very high detection fractions and achieve similar depths as the \BlueTides sample. We therefore do not consider deeper surveys such as JADES-Deep, which will discover fainter galaxies. 
However, these surveys will be important for detecting even the brightest higher-$z$ galaxies ($z\gtrsim10$), rare objects which can be predicted in greater detail by a larger volume simulation such as FLARES \citep[e.g.][]{Lovell2020,Wilkins2022}.

\subsubsection{Methodology for Survey Predictions}

We make predictions for these surveys by adding their expected noise to our mock galaxy images, from their quoted 10$\sigma$ limits or, if unavailable, estimating the noise from the exposure time as described in Section \ref{sec:Photometry}. As above, we then perform photometry on these mock images, determining whether each high-$z$ galaxy is detected. We require a galaxy to be detected in at least 2 filters within the survey for it to be classified as `detected' in the survey overall. 
We note that we do not simulate a full mock survey image using a light-cone, and are simply investigating whether the individual galaxies could be detected in a survey image of equivalent depth, ignoring any effects such as foreground contamination.
However, when considering the survey areas, we use the true positions of each galaxy from the \BlueTides volume in combination with the image cut-out, so in this way we are accurately representing the galaxy distribution at each redshift that could be probed by each survey.

The number of galaxies in each survey area will depend on its placement on the sky relative to the overall galaxy distribution, and particularly galaxy overdensities.
To account for such cosmic variance, we consider sub-cubes of the full \BlueTides box with volume $V = \textrm{Survey Width} \times \textrm{Survey Height} \times D_C(z'-0.5\textrm{ to }z'+0.5)$. The co-moving radial distance gives the appropriate redshift depth corresponding to $z'-0.5<z<z'+0.5$, as opposed to the full $z=z'$ \BlueTides snapshot box which has a depth of $400/h$ cMpc (see Section \ref{Sec:GalaxySample}).
We consider the number of non-overlapping sub-cubes that can fit in the \BlueTides volume, each corresponding to an independent, single realisation of the area covered by the given survey.
We note that each of the \textit{JWST} surveys may have been chosen to cover a specific region of known density, however our calculations assume no preference on targeted environment. Note that we also do not consider the effect of gravitational lensing.

\subsubsection{Method Verification with \textit{HST}}

To verify that our mock images and photometry are working as expected, we first make predictions for an existing \textit{HST} survey.
We consider the CANDELS EGS field \citep{grogin_2011}, chosen as it is a medium-depth survey, has easily accessible publicly available catalogues, and has an almost constant depth across the field which allows for a simpler comparison.
The EGS field has an area of $\sim 7' \times 26'$, with exposures of 1900s and 3600s in the WFC3 F125W and F160W filters, respectively. This corresponds to approximate 5$\sigma$ point source sensitivities of 27.2 and 27.0 mag \citep{grogin_2011}, which we use in the simulation as our noise estimate. We also resample the images by a factor of 2, to the native pixel scale of $0\farcs13$.

We search the CANDELS catalogue from \citet{Stefanon2017} for galaxies with $6.5<z_{\rm{phot, median}}<7.5$. Given the uncertainty in the photometric redshifts, we require each of the various photometric redshift measurements considered in  \citet{Stefanon2017} to be $z_{\rm{phot}}>3$, to be confident that the selected galaxies are not low-$z$ interlopers. There are 2 such galaxies in the EGS field that have magnitudes in the \BlueTides magnitude range, $24<M_{J/H}<26.75$ mag. This is a rough approximation of the number of galaxies in $6.5<z<7.5$ in the true HST survey, as these are not spectroscopically-confirmed galaxies.

We run our photometry algorithm on the mock F125W and F160W images. For 178 of the sub-cubes, no galaxies are detected.  In 46 sub-cubes one galaxy is detected, 16 sub-cubes contain two detected galaxies, in 6 sub-cubes three galaxies are detected, and one sub-cube contains four detected galaxies. This is broadly consistent with the CANDELS EGS which detected approximately two galaxies. We are therefore confident that our method works as expected and is a reasonable estimate for true observations.

\subsubsection{Predictions for \textit{JWST}}
\label{Sec:SurveyResults}

\begin{figure*}
\begin{center}
\includegraphics[scale=0.9]{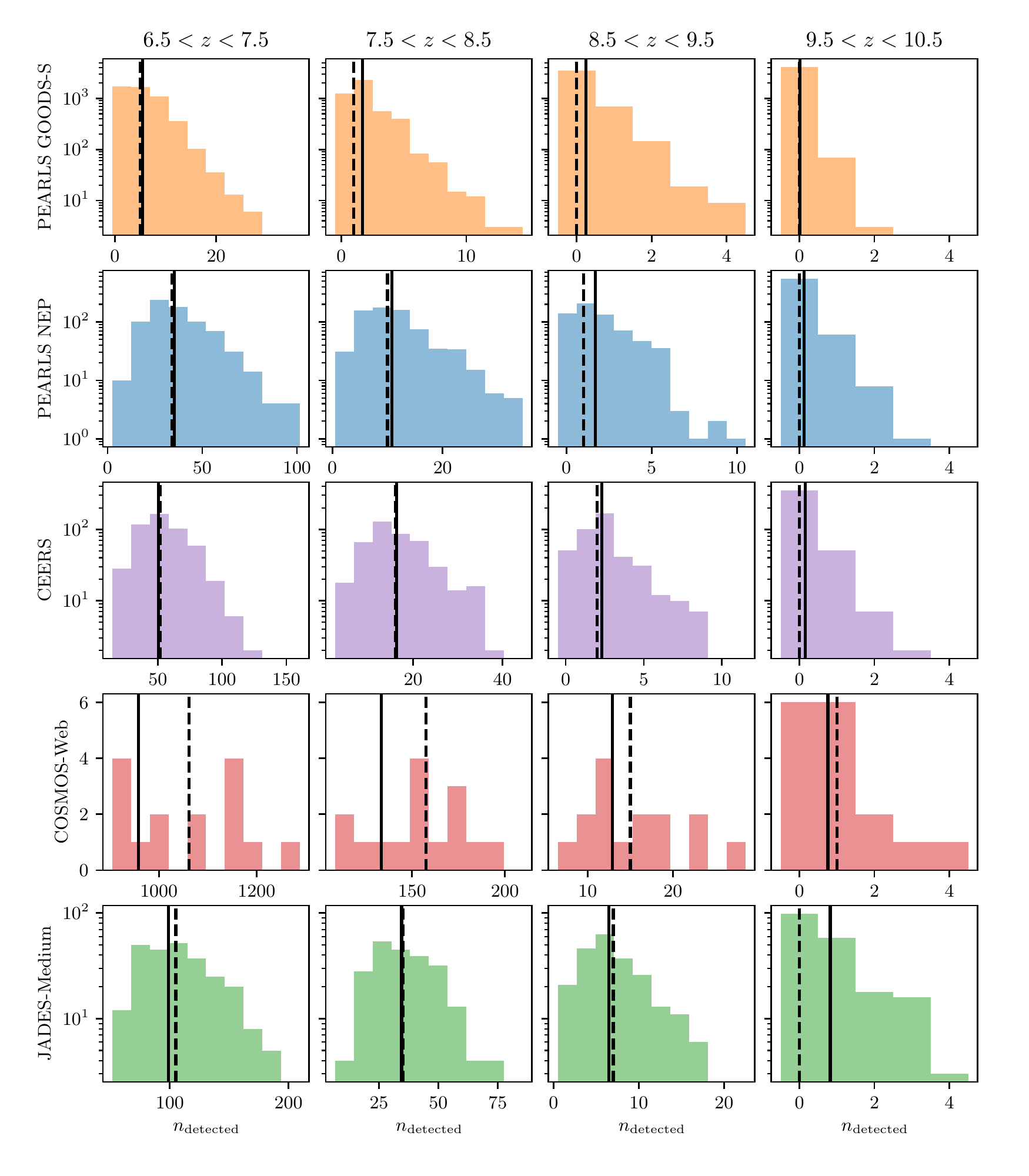}
\caption{Histograms of the number of galaxies successfully detected in each \BlueTides sub-cube, for each redshift range and survey; i.e. the vertical axis shows the number of sub-cubes (i.e. survey realisations) in which the specified number of galaxies is detected. Each sub-cube is a non-overlapping region within \BlueTides with volume corresponding to the survey volume; the size of the survey determines the number of sub-cubes possible.
The solid black lines show the mean number of galaxies detected in the sub-cubes, while the dashed black lines show the median, which are given in Table \ref{table:Detections}.
Note that these are the number of detected galaxies above the completeness limit of \BlueTidesns: $M_{\textrm{1500\AA}}<-20.1$ mag for $z=7$ and $M_{\textrm{1500\AA}}<-19.9$ mag for $z>7$. }
\label{Fig:SurveyNumberCounts}
\end{center}
\end{figure*}

\begin{table*}
\caption{The \BlueTides predictions for the number of detected galaxies $n_{\textrm{detected}}$ for each survey, at each redshift range.
We give the mean and median $n_{\textrm{detected}}$, the average detection fraction $n_{\textrm{detected}}/n_{\textrm{total}}$, and the minimum and maximum $n_{\textrm{detected}}$ from the various sub-cube realisations of each survey, due to cosmic variance. These are the number of detected galaxies above the completeness limit of \BlueTidesns: $M_{\textrm{1500\AA}}<-20.1$ mag for $z=7$ and $M_{\textrm{1500\AA}}<-19.9$ mag for $z>7$.}
\begin{tabular}{lcrrcc}
\hline 
Survey & Redshift & Mean $n_{\rm{detected}}$ & Median $n_{\rm{detected}}$ & Average $n_{\rm{detected}}/n_{\rm{total}}$ & (Min, Max) $n_{\rm{detected}}$\\ 
\hline
PEARLS GOODS-S & $6.5<z<7.5$ & 5.48 & 5 & 0.999 & (0,37) \\
& $7.5<z<8.5$ & 1.68 & 1 & 0.908 & (0,15) \\
& $8.5<z<9.5$ & 0.24 & 0 & 0.543 & (0,4) \\
& $9.5<z<10.5$ & 0.02 & 0 & 0.193 & (0,2) \\
& $10.5<z<11.5$ & 0.00 & 0 & 0.144 & (0,1) \\
& $11.5<z<12.5$ & 0.00 & 0 & 0.000 & (0,0) \\
\hline
PEARLS NEP & $6.5<z<7.5$ & 35.36 & 34 & 0.998 & (3,102) \\
& $7.5<z<8.5$ & 10.81 & 10 & 0.913 & (1,35) \\
& $8.5<z<9.5$ & 1.69 & 1 & 0.575 & (0,11) \\
& $9.5<z<10.5$ & 0.12 & 0 & 0.210 & (0,3) \\
& $10.5<z<11.5$ & 0.02 & 0 & 0.153 & (0,1) \\
& $11.5<z<12.5$ & 0.00 & 0 & 0.083 & (0,0) \\
\hline
CEERS & $6.5<z<7.5$ & 50.63 & 52 & 0.995 & (15,161) \\
& $7.5<z<8.5$ & 16.33 & 16 & 0.902 & (3,45) \\
& $8.5<z<9.5$ & 2.31 & 2 & 0.547 & (0,12) \\
& $9.5<z<10.5$ & 0.16 & 0 & 0.187 & (0,3) \\
& $10.5<z<11.5$ & 0.03 & 0 & 0.139 & (0,1) \\
& $11.5<z<12.5$ & 0.00 & 0 & 0.000 & (0,0) \\
\hline
COSMOS-Web & $6.5<z<7.5$ & 958.44 & 1061.5 & 0.841 & (905,1289) \\
& $7.5<z<8.5$ & 133.52 & 157.5 & 0.377 & (109,210) \\
& $8.5<z<9.5$ & 12.92 & 15 & 0.155 & (7,29) \\
& $9.5<z<10.5$ & 0.76 & 1 & 0.052 & (0,4) \\
& $10.5<z<11.5$ & 0.04 & 0 & 0.010 & (0,1) \\
& $11.5<z<12.5$ & 0.00 & 0 & 0.000 & (0,0) \\
\hline
JADES-Medium & $6.5<z<7.5$ & 98.81 & 105 & 1.000 & (52,210) \\
& $7.5<z<8.5$ & 34.54 & 35 & 0.997 & (7,86) \\
& $8.5<z<9.5$ & 6.51 & 7 & 0.830 & (1,23) \\
& $9.5<z<10.5$ & 0.82 & 0 & 0.453 & (0,5) \\
& $10.5<z<11.5$ & 0.11 & 0 & 0.361 & (0,2) \\
& $11.5<z<12.5$ & 0.00 & 0 & 0.125 & (0,0) \\
\hline
\end{tabular} 
\label{table:Detections}
\end{table*}

\begin{figure*}
\begin{center}
\includegraphics[scale=0.9]{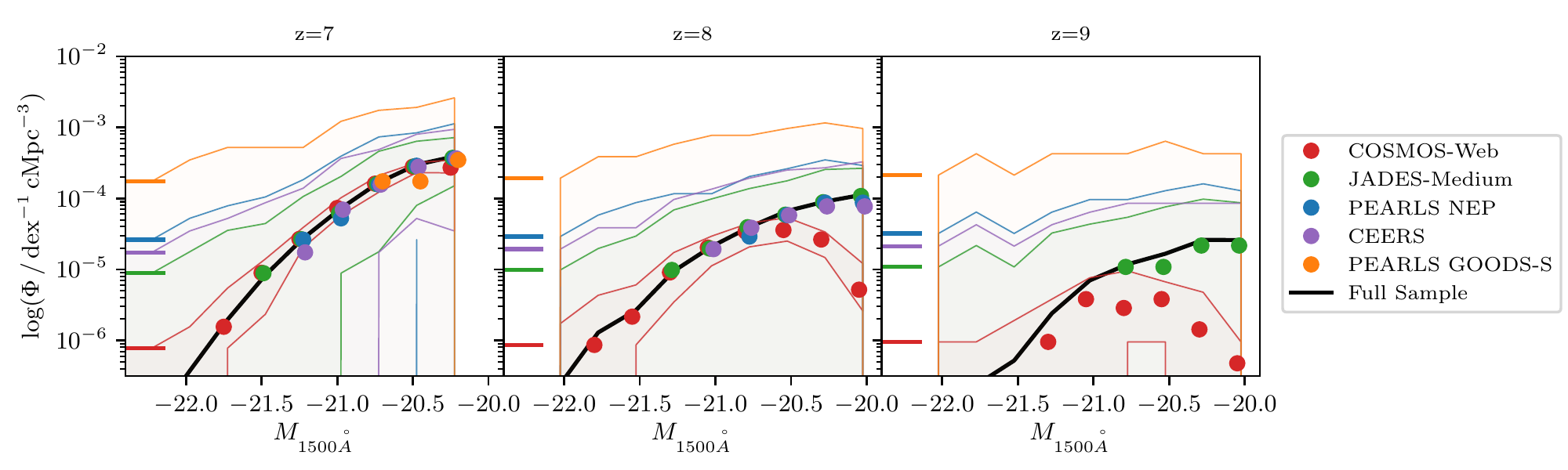}
\caption{The luminosity function predicted to be measured by each survey (see legend), for redshifts $z=7$, 8, and 9. The points show the median number of galaxies detected in the magnitude bin, for each of the sub-cubes representing each survey; the shaded regions of the same colour cover the minimum and maximum number of detected galaxies in the sub-cubes.
Bins have a width of 0.25 mag. There is a very small horizontal offset for the points of each survey to improve readability.
The black line is the total luminosity function of \BlueTides galaxies.
The coloured lines on the left of each panel show where the number density corresponds to 1 galaxy/dex being found in the survey area.
}
\label{SurveyLFs}
\end{center}
\end{figure*}

We now show our predictions for the number of high-$z$ galaxies that will be detected in each of the \textit{JWST} surveys.
For the various redshifts and surveys, histograms of the number of detected galaxies in each of the \BlueTides sub-cubes with the volume of the survey (i.e. survey realisation), are shown in Figure \ref{Fig:SurveyNumberCounts}. The mean, median, minimum and maximum number of detected galaxies $n_{\textrm{detected}}$ in each set of sub-cubes are listed in Table \ref{table:Detections}.
For a best estimate of the average fraction of galaxies that are successfully detected in each survey, $n_{\textrm{detected}}/n_{\textrm{total}}$, we consider the full sample of galaxies in the \BlueTides volume at the redshift snapshot, and not the individual sub-cubes, for the largest possible sample size.
These values are also listed in Table \ref{table:Detections}.
Note that we only consider the number of detected galaxies above the completeness limit of \BlueTidesns: $M_{\textrm{1500\AA}}<-20.1$ mag for $z=7$, and $M_{\textrm{1500\AA}}<-19.9$ mag for $z>7$.

We predict that the COSMOS-Web survey will detect the most high-$z$ galaxies: $\sim1000$ galaxies in $6.5<z<7.5$. The detection fraction of this shallow survey of 84\% is lower than those for the deeper surveys, however the very large volume of COSMOS-Web still results in the most detected high-$z$ objects. 

In Figure \ref{SurveyLFs} we show the luminosity function that would be obtained by each survey, with the points showing the median number density of detected galaxies obtained over the various sub-cubes, and the shaded regions covering the minimum and maximum values.
Surveys with a large area such as COSMOS-Web will very precisely measure the true sample luminosity function, with each sub-cube resulting in a similar measurement. This is due to the large area, which covers a wide range of galaxy environments. Surveys with smaller areas could measure a wider range of luminosity functions, depending on the region they sample---i.e. due to cosmic variance.
The larger area of COSMOS-Web will also allow the detection of bright high-$z$ galaxies that are not statistically expected in the smaller surveys due to their low number densities. However, at higher-$z$ the COSMOS-Web luminosity function will become incomplete at the fainter magnitudes that the deeper surveys will be able to cover.

The large depth of the JADES-Medium survey will result in detections of $\sim100$ galaxies in $6.5<z<7.5$, with detection fractions of almost 100\% of the $M_{\textrm{1500\AA}}<-20.1$ mag $6.5<z<7.5$ galaxies and $M_{\textrm{1500\AA}}<-19.9$ mag $7.5<z<8.5$ galaxies in its field of view. At these redshifts the number of galaxies detected is limited only by the number of objects in the field of view; hence the smaller but deeper JADES-Deep survey would offer no improvements on the detected number. However, additional depth would improve the higher-$z$ detection fraction, as well as detecting fainter galaxies that we cannot probe due to the limited resolution of \BlueTidesns.
JADES-Medium is likely to result in accurate measurements of the fainter-end of the luminosity function that the shallow depth of COSMOS-Web cannot reach, particularly at $z>7$ (Figure \ref{SurveyLFs}). 

The more modest PEARLS GOODS-S and NEP fields have similar depths, with the NEP covering $\sim6$ times the area. On average we predict that they will detect $\sim5$ and $\sim35$ galaxies in $6.5<z<7.5$ respectively, with $M_{\textrm{1500\AA}}<-20.1$ mag.
The CEERS field has an equivalent minimum depth as the NEP field, covering a survey area $\sim1.5$ times larger. This results in predicted galaxy counts that are $\sim1.5$ times larger than that from the NEP, $\sim51$ galaxies in $6.5<z<7.5$. Note, however, that the CEERS program will observe some pointings with larger exposure times, and so our predictions are lower limits for the number of $M_{\textrm{1500\AA}}\lesssim-20$ mag galaxies that this survey will detect at $z\gtrsim7$.

Due to cosmic variance, there is a large range in expected galaxy counts for these smaller fields (Figure \ref{Fig:SurveyNumberCounts}). From Figure \ref{SurveyLFs}, we see that the PEARLS NEP and CEERS fields are expected to give reasonable estimates of the overall luminosity function from $M_{\textrm{1500\AA}}\simeq-21$ mag to our \BlueTides completeness limit of $M_{\textrm{1500\AA}}\simeq-20$ mag, at $z=7$ and $z=8$. The \BlueTides luminosity function drops below the number density corresponding to 1 galaxy/dex being detected in these surveys at $z=9$. This occurs at $z>7$ for the smaller PEARLS GOODS-S field. In this case, the number of detected galaxies will be highly dependent on the sampled environment. If galaxies are detected, this is likely due to the survey observing an over-dense region, and so the measured luminosity function will be artificially higher than the overall galaxy luminosity function; this is particularly the case for the PEARLS GOODS-S field, as seen from the shaded region in Figure \ref{SurveyLFs}. 
Thus the CEERS and PEARLS NEP surveys are not expected to provide an accurate measure of the $M_{\textrm{1500\AA}}<-20$ mag luminosity function at $z>8$, or $z>7$ for PEARLS GOODS-S. 
Care must be made with smaller volume surveys when interpreting measured luminosity functions, due to cosmic variance

\section{Discussion}
\label{Sec:Discussion}
\subsection{Comparison to Existing Predictions}
The FLARES zoom-in simulation suite has larger volume and resolution than \BlueTidesns, containing a large number of very high-$z$ galaxies and allowing for detailed $z>10$ predictions \citep{Wilkins2022}. In FLARES, \citet{Wilkins2022} estimate that in the PEARLS NEP field there will be $\sim20$ $z\geq8$ galaxies, with $\sim70$ in CEERS and $\sim350$ in COSMOS-Web. We predict fewer galaxies in all cases, with a median of $\sim11$, 18 and 174 galaxies at $z>7.5$ in each survey respectively. 
However, we are limited to $M_{\textrm{1500\AA}}<-20$ mag, and both FLARES and the observations will have fainter objects that can be detected, so we expect to predict fewer galaxies. We also expect to predict fewer galaxies as we consider a full photometric extraction, as opposed to using the luminosity function and a 100\% completeness cut at the point-source magnitude as in \citet{Wilkins2022}.
It is interesting to note that our predictions are approximately half of those of FLARES for the NEP and COSMOS-Web, but we predict almost 4 times fewer galaxies in CEERS; this may be because we assume the shallowest depth which covers the full CEERS field, not considering that some pointings have longer exposure times.

\citet{Behroozi2020} use the \textsc{UniverseMachine} model to provide an estimate of the number of galaxies per square arcminute above $M_{\textrm{1500\AA}}<-20$ mag, in an integer redshift range. Using this and the areas of the five surveys from Table \ref{Table:Surveys}, and assuming a 100\% detection fraction, we convert these to predictions for JADES-Medium, COSMOS-Web, CEERS, PEARLS NEP and PEARLS GOODS-S of 76, 864, 39, 26 and 4 galaxies at $z=7$, respectively. These predictions are consistent with our expected range as estimated from the various survey sub-cubes (Table \ref{table:Detections}). At $z=10$, the \citet{Behroozi2020} model corresponds to predictions of 5, 54, 2.4, 1.6 and 0.2 galaxies in the five surveys. We note that these are significantly more than our predictions, which all have a mean of $<1$ and with a maximum of 5 $z=10$ galaxies detected in any of the survey realisations. However, to account for the fact that these \citet{Behroozi2020} predictions assume that every galaxy above $M_{\textrm{1500\AA}}<-20$ mag is detected, we multiply these predictions by our detection fractions from Table \ref{table:Detections}. We find that, as high-$z$ galaxies are difficult to recover from the photometry, the \citet{Behroozi2020} predictions are reduced to only 2, 3, 0.4, 0.3 and 0.03 galaxies in the five surveys at $z=10$; these are consistent with our \BlueTides predictions.

Similarly, using the galaxy number density from the Illustris-TNG simulation \citep[figure 16,][]{Vogelsberger2020}, the survey areas, and assuming a 100\% detection fraction, we predict that CEERS, JADES-Medium, and COSMOS-Web will
detect $\sim13$, 26 and 301 galaxies at $z=8$ with $M_{\textrm{1500\AA}}<-19.9$ mag, and $\sim30$, 59 and 673 at $z=7$ with $M_{\textrm{1500\AA}}<-20.1$ mag, respectively. These are broadly consistent with our estimates and generally lie within our expectations due to cosmic variance. However, this predicts only $\sim70$\% of the number of $z=7$ galaxies we expect in COSMOS-Web from \BlueTidesns, yet  approximately twice as many at $z=8$. These predictions assume 100\% detectability above the magnitude threshold; if we multiply by our detection fractions in Table \ref{table:Detections}, \citet{Vogelsberger2020} would predict that COSMOS-Web
will detect $\sim113$ galaxies at $z=8$ with $M_{\textrm{1500\AA}}<-19.9$ mag, and $\sim566$ galaxies at $z=7$ with $M_{\textrm{1500\AA}}<-20.1$ mag; $\sim70$\% and $\sim50$\% of our predictions, which is more consistent.

\subsection{Limitations}

Our approach to creating mock telescope images to predict the number of detectable high-$z$ galaxies includes several limitations. 

For \textit{JWST}, our noise estimates all assumed a background of 1.2 times the minimum zodiacal light background at RA = 17:26:44, Dec = -73:19:56 on June 19, 2019, as predicted by pre-flight data \citep{NIRCamImaging}. The true depth of an image will depend on the position of the pointing on the sky, the date of observation, and the readout mode. 
For a specific example, the PEARLS NEP field is in a dark region of sky, and so our assumed background is likely to be higher than the observed background.
In addition, the \textit{JWST} science instrument commissioning has predicted significantly better $10\sigma$ sensitivities than the pre-launch values used throughout this work \citep{Rigby2022}, which would cause our predictions to be underestimates of the completeness and expected number of detectable galaxies.

Our images are cut-outs of individual galaxies at specific redshift snapshots.
We do not make light-cones of galaxies across a range of redshifts, simulating a true survey image as in \citet{Williams2018} and \citet{Drakos2022}, for example.
Thus, we cannot consider the effects of blending, with nearby foreground galaxies or stars complicating the detection of high-$z$ galaxies. In addition, our method implicitly assumes that all of the galaxies that are successfully detected are indeed at the specified redshift. With true observations, contaminating low-$z$ objects may be misidentified as high-$z$ galaxies, or alternatively true high-$z$ galaxies may be miscategorised. 
We do not consider this potential contamination, which was investigated for \textit{JWST} in detail in \citet{Hainline2020}, for example.
Instead, with our approach we simply study whether high-$z$ galaxies could be successfully detected in a specific image, assuming that they are isolated in the image and are correctly identified.

A key limitation of this work is that we are restricted to studying only galaxies with $M_{\textrm{1500\AA}}\lesssim-20$ mag, due to the resolution of the simulation.
Unfortunately this limits our predictions to the brightest high-$z$ galaxies, while one of the key features of \textit{JWST} is its exquisite sensitivity, which will allow it to discover faint objects in the early Universe.
However, the large volume of \BlueTides makes it an ideal simulation for studying bright, rare,  high-$z$ galaxies. 
This work has thus provided detailed predictions for these bright galaxies that are limited in smaller simulations. In addition, the large volume of \BlueTides allows us to study the effects of cosmic variance on the predicted galaxy number counts.

\section{Conclusions}
\label{Sec:Conclusion}

In this work we introduce the \BlueTides Mock Image Catalogue, a publicly available catalogue of mock images of $\sim100,000$ $M_{UV}\simeq-22.5$ to $-19.6$ mag galaxies in the \BlueTides hydrodynamical simulation at $z=7$--12. We create mock images with the \textit{James Webb}, \textit{Hubble}, \textit{Euclid} and \textit{Roman Space Telescopes}, as well as VISTA and Subaru.

These images are created from the stellar particle distribution of \BlueTides galaxies, as determined through the detailed hydrodynamics of the cosmological simulation.  Each star particle is assigned an SED based on its age and metallicity, and nebular continuum, line emission, and dust attenuation from the ISM and a birth cloud are also modelled. Images are created with the pixel scales of the various instruments, with fluxes taken from a convolution of the SEDs with the various filter transmission curves (see Appendix \ref{Appendix}). These images are convolved with model PSFs, to produce realistic estimates of what true images with these telescopes would look like. 
The available images are $6\times6$ and $10\times10$ kpc snapshots around each galaxy, and not a full mock light-cone or instrument field of view.
We note that the images available in the catalogue have no noise, so that they can be adapted for specific use cases.

We use these mock images to make detailed predictions for photometric surveys with the various telescopes. To do this we add appropriate noise to the images, and then run a \textsc{Photutils} photometric source extraction algorithm to determine whether the galaxies would be successfully detected in the given exposures.

To compare the various telescopes we perform photometry on mock 10ks $Y$-band images of $z=7$ galaxies from VISTA, Subaru, \textit{HST}, \textit{JWST} and \textit{Roman}.
We predict the highest detection fractions from \textit{JWST}, which becomes 95\% complete at $M_{\textrm{1500\AA}}=-20.2$ mag, at least 2.5 magnitudes fainter than VISTA and Subaru, 1.1 magnitudes fainter than \textit{HST}, and 0.9 magnitudes fainter than \textit{Roman}. This highlights the remarkable capabilities of \textit{JWST} for detecting high-$z$ galaxies.

We then consider various observing strategies with \textit{JWST}, adding noise corresponding to a range of exposure times, for each of the simulated NIRCam and MIRI filters. We find that 10ks observations become 95\% complete at an average of 26.7, 26.9, 27.1, 27.4, 27.4, 26.9, 27.0, 25.7, and 24.9 mag for F115W, F150W, F200W, F277W, F356W, F410M, F444W, F560W, and F770W, respectively, with a variation of $\pm0.1$ mag between the various redshifts for most filters. The F277W
and F356W filters can detect the faintest galaxies in a fixed exposure time, with the MIRI filters the least sensitive.
Our 95\% completeness limits are 1.9 to 0.5 magnitudes brighter than the predicted 10$\sigma$ point source sensitivities, as our galaxies are extended and have realistic structures. We also find that increasing the exposure time from 1 to 10ks is expected to improve the NIRCam detection depth by approximately 1.25--1.6 magnitudes.

We then make predictions for five upcoming \textit{JWST} Cycle 1 surveys: JADES-Medium, COSMOS-Web, CEERS, PEARLS NEP and PEARLS GOODS-S. We add realistic noise estimates based on the survey depths in each filter, and require a galaxy to be detected in at least two of the filters for it to be classified as successfully detected. To consider the effects of cosmic variance, we bin the full \BlueTides cube into sub-cubes with volume equal to the survey volume, with each sub-cube corresponding to an independent realisation of the survey.

We predict that the COSMOS-Web survey will detect the most high-$z$ $M_{\textrm{1500\AA}}<-20$ mag galaxies, with an average of $\sim1000$ galaxies expected in $6.5<z<7.5$.
With its large survey area, COSMOS-Web will detect bright high-$z$ galaxies that are too rare to be found in smaller surveys, and will precisely measure the bright end of the galaxy luminosity function as the large area reduces the effects of cosmic variance.
We predict that JADES-Medium will detect $M_{\textrm{1500\AA}}\lesssim-20$ mag galaxies with $\sim100$\% success rates at $z<8.5$, detecting $\sim100$ galaxies with $M_{\textrm{1500\AA}}\leq-20.1$ mag in $6.5<z<7.5$. This deeper survey will obtain accurate measurements of the fainter-end of the luminosity function that the shallow depth of COSMOS-Web cannot reach.
We predict that the PEARLS GOODS-S and NEP fields will detect $\sim5$ and $\sim35$ $M_{\textrm{1500\AA}}\lesssim-20$ mag galaxies in $6.5<z<7.5$ respectively, with CEERS detecting $\sim50$ such galaxies.
These smaller volume surveys are highly subject to cosmic variance, with a wide range in expected number counts and measured luminosity functions depending on the region they sample.

\BlueTides is an ideal simulation for studying bright, rare galaxies in the early Universe. Due to the resolution of the simulation however, we are limited to only studying galaxies above the completeness limit of $M_{\textrm{1500\AA}}=-20.1$ mag at $z=7$, and $M_{\textrm{1500\AA}}=-19.9$ mag at $z>7$. We have therefore focused only on shallower surveys and made predictions for these bright galaxies. Overall, \textit{JWST} is expected to discover many fainter and higher-$z$ galaxies that this simulation cannot accurately capture.

Our approach of applying photometry to mock images of these \BlueTides galaxies gives us detailed predictions for realistic galaxies, as opposed to the common approaches of using a luminosity function and magnitude cut, or applying photometry to modelled galaxies with S\'ersic profiles. As well as our \textit{JWST} investigation, this approach using the \BlueTides Mock Image Catalogue can be used in future work for \textit{Roman} and \textit{Euclid} survey predictions, for example. The images can also be used for studying the physical properties of galaxies within the \BlueTides simulation, such as their sizes and morphologies, and determining how successfully these properties can be measured with various instruments. Overall, this publicly available \BlueTides Mock Image Catalogue will be a useful tool for the community.

\section*{Data Availability}
The \BlueTides Mock Image Catalogue is publicly available as a High Level Science Product (HLSP) via the Mikulski Archive for Space Telescopes (MAST) archive at \url{https://doi.org/10.17909/er09-4527}. Example codes for using this catalogue can be found at \url{https://github.com/madelinemarshall/BlueTidesMockImageCatalogue}.
Data of the \textsc{BlueTides} simulation is available at \url{http://bluetides.psc.edu}.
Other data generated in this work will be shared on reasonable request to the corresponding author.

\section*{Acknowledgements}
We sincerely thank Claire Murray, Scott Fleming, Greg Snyder, and the MAST team for their work making our catalogue available as a HLSP. We are very grateful to the anonymous referee for their help improving this manuscript.

The \BlueTides simulation was run on the BlueWaters facility at the National Center for Supercomputing Applications.
Part of this work was performed on the OzSTAR national facility at Swinburne University of Technology, which is funded by Swinburne University of Technology and the National Collaborative Research Infrastructure Strategy (NCRIS).
MAM acknowledges the support of a National Research Council of Canada Plaskett Fellowship, and the Australian Research Council Centre of Excellence for All Sky Astrophysics in 3 Dimensions (ASTRO 3D), through project number CE170100013.
TDM acknowledges funding from NSF ACI-1614853, NSF AST-1517593, NSF AST-1616168 and NASA ATP 19-ATP19-0084 and 80NSSC20K0519,ATP.
TDM and RAC also acknowledge ATP 80NSSC18K101 and NASA ATP 17-0123.
The Cosmic Dawn Center (DAWN) is funded by the Danish National Research Foundation under grant No. 140.

This paper made use of Python packages and software 
AstroPy \citep{Astropy2013},
BigFile \citep{Feng2017},
FLARE \citep{FLARE},
Matplotlib \citep{Matplotlib2007},
NumPy \citep{Numpy2011},
Pandas \citep{reback2020pandas}, 
Photutils \citep{photutils},
SciPy \citep{2020SciPy-NMeth},  
and SynthObs \citep{SynthObs}.
This paper also makes use of version 17.00 of \textsc{Cloudy}, last described by \citet{Ferland2017}, and version 2.2.1 of the Binary Population and Spectral Population Synthesis (BPASS) model \citep{Stanway2018}.

\bibliographystyle{mnras}
\bibliography{TelescopePredictionsPaper.bib} 


\appendix
\section{Available Filters}
\label{Appendix}
\begin{figure}
\begin{center}
\includegraphics[scale=0.75]{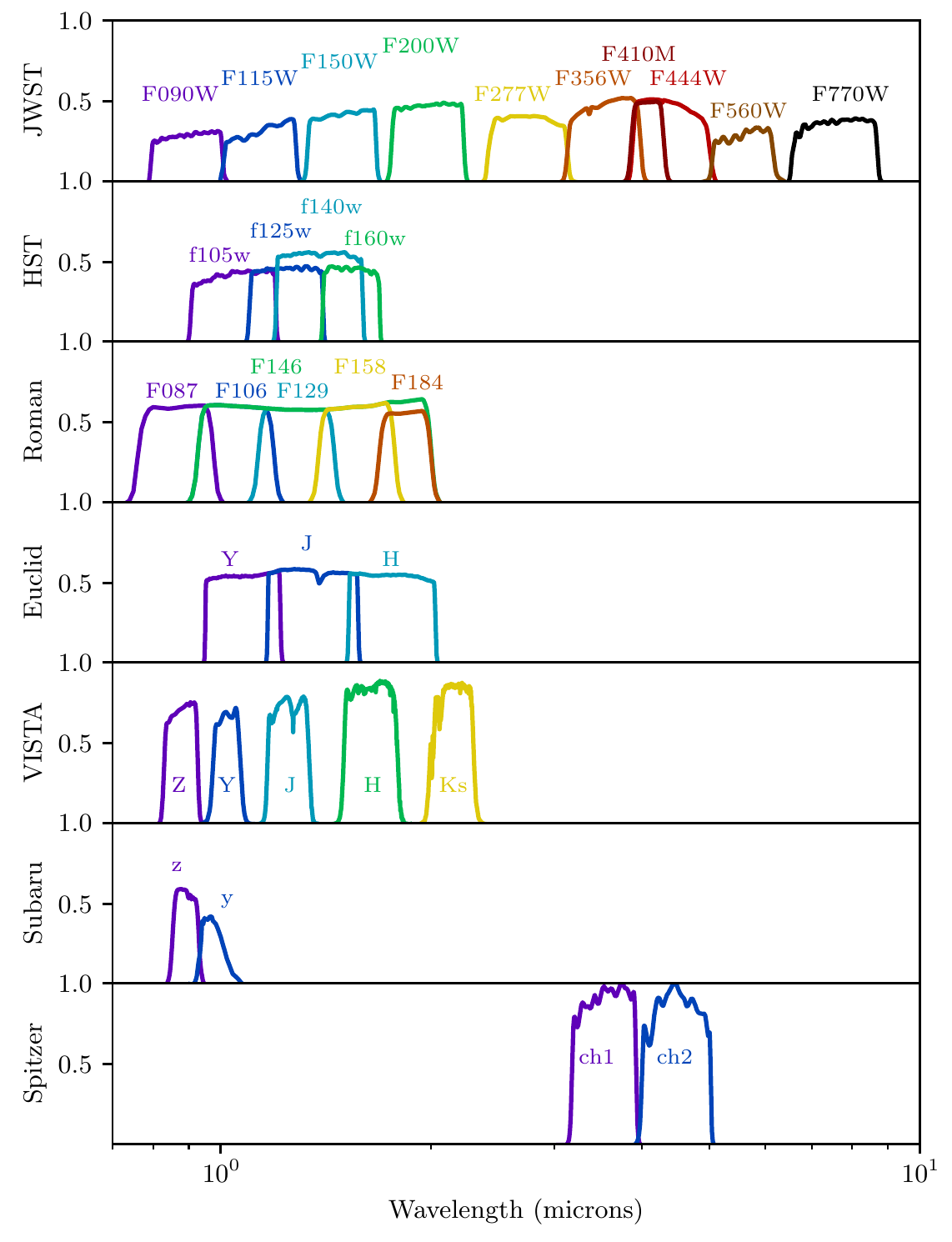}
\caption{The filter transmission curves used in the simulations for each telescope. Note that we provide only the \textit{Spitzer} fluxes and not images, as galaxies are unresolved.}
\label{Filters}
\end{center}
\end{figure}

Here we provide a visual of the available filters in the \BlueTides mock image catalogue (Figure \ref{Filters}).

\bsp	
\label{lastpage}
\end{document}